\documentclass{article}
\usepackage{graphicx}
\usepackage{amsmath}
\usepackage{amsfonts}
\usepackage{amssymb}
\newtheorem{theorem}{Theorem}
\newtheorem{acknowledgement}[theorem]{Acknowledgement}

\begin{document}

\title{{\Large Spin equation and its solutions}}
\author{V.G. Bagrov\thanks{On leave from Tomsk State University and Tomsk Institute of
High Current Electronics, Russia, e-mail: bagrov@phys.tsu.ru}, D.M.
Gitman\thanks{e-mail: gitman@dfn.if.usp.br}, M.C. Baldiotti\thanks{e-mail:
baldiott@fma.if.usp.br}, and A.D. Levin\thanks{Dexter Research Center, USA;
e-mail: SLevin@dexterresearch.com},\\ \\Instituto de F\'{\i}sica, Universidade de S\~{a}o Paulo,\\Caixa Postal 66318-CEP, 05315-970 S\~{a}o Paulo, S.P., Brazil}
\maketitle
\begin{abstract}
The aim of the present article is to study in detail the so-called spin
equation (SE) and present both the methods of generating new solution and a
new set of exact solutions. We recall that the SE\ with a real external field
can be treated as a reduction of the Pauli equation to the ($0+1$)-dimensional
case. Two-level systems can be described by an SE with a particular form of
the external field. In this article, we also consider associated equations
that are equivalent or (in one way or another) related to the SE. We describe
the general solution of the SE and solve the inverse problem for this
equation. We construct the evolution operator for the SE and consider methods
of generating new sets of exact solutions. Finally, we find a new set of exact
solutions of the SE.
\end{abstract}

\section{Introduction}

We refer as the spin equation (SE) to the following set of two ordinary linear
differential equations of first order for the functions $v_{1}\left(
t\right)  $ and $v_{2}\left(  t\right)  $:%
\begin{equation}
i\dot{V}=\left(  \mathbf{\sigma F}\right)  V{\LARGE \,},\;V=\left(
\begin{array}
[c]{c}%
v_{1}\left(  t\right) \\
v_{2}\left(  t\right)
\end{array}
\right)  \,,\;\dot{V}=dV/dt\,. \label{1.1}%
\end{equation}
Here, $\mathbf{\sigma}=\left(  \sigma_{1},\sigma_{2},\sigma_{3}\right)  $ are
the Pauli matrices, and $\mathbf{F}$ is a given time-dependent (in general,
complex) vector,
\begin{equation}
\mathbf{F}=\left(  F_{k},\;k=1,2,3\right)  ,\;F_{k}=F_{k}\left(  t\right)  \,.
\label{1.2}%
\end{equation}
In what follows, the column $V$ and the vector $\mathbf{F}$ are called the
spinor field and the external field, respectively.

The SE\ with a real external field can be treated as a reduction of the Pauli
equation \cite{Pau27} to the ($0+1$)-dimensional case. Such an equation is
used to describe a (frozen in space)\ spin-$1/2$ particle of magnetic momentum
$\mu$, immersed in a magnetic field $\mathbf{B}$ (in this case, $\mathbf{F}%
=-\mu\mathbf{B}$), and\textbf{ }was intensely studied in connection with the
problem of magnetic resonance \cite{RabRaS54,BloSi40}. Besides, complex
quantum systems with a discrete energy spectrum can be placed in a special
dynamic configuration in which only two stationary states are important. In
those cases, it is possible to reduce the Hilbert space of the system to a
two-dimensional space. Such two-level systems can also be described by the SE.
The SE with an external field of the form $\mathbf{F}=\left(  F_{1}%
,0,F_{3}\right)  ,$ $F_{1}=\epsilon,$ where $\epsilon$ is a constant,
describes two-level systems with unperturbed energy levels $\pm\epsilon$
($F_{3}\equiv0$) submitted to an external time-dependent interaction
$F_{3}(t)$, inducing a transition between the unperturbed eigenstates.
Two-level systems possess a wide range of applications, for example, the
semi-classical theory of laser beams \cite{Nus73}, the absorption resonance
and nuclear induction experiments \cite{RabRaS54}, the behavior of a molecule
in a cavity immersed in electric or magnetic fields \cite{FeyVe57}, and so on.
Recently, this subject attracts even more attention due to its relation to
quantum computation \cite{QuantComp}, where the state of each bit of
conventional computation is permitted to be any quantum-mechanical state of a
\textit{qubit} (quantum bit), which can be treated as a two-level system. The
SE with complex external fields describes a possible damping of two-level
systems. There exist various equations that are equivalent, or (in a sense)
related, to the SE. For example, the well-known top equation, which appears in
the gyroscope theory, in the theory of precession of a classical gyromagnet in
a magnetic field (see \cite{FeyVe57}), and so on. The SE with an external
field in which $F_{s}\left(  t\right)  ,\;s=1,2$ are purely imaginary and
$F_{3}$ is constant is a degenerate case of the Zakharov--Shabat equation,
which plays an important role in the soliton theory \cite{ZakSh84}. It turns
out that two-level systems present a convenient object for illustrating and
applying the geometrical phase method \cite{ShaWi89}. For periodic, or
quasiperiodic, external fields, the equations of a two-level system have been
studied by many authors using different approximation methods, e.g.,
perturbative expansions \cite{BarCo02}, see also \cite{Bar00}. When the
external field $\mathbf{F}$ is not periodic, or quasiperiodic, there exists no
regular approach to finding exact solutions of the SE. One ought to stress
that exact solutions are very important in view of the numerous physical
application of the SE. Since in the general case the problem of finding such
solutions is sufficiently involved, it has been solved only in a few
particular cases.

The first exact solution of the SE was found by Rabi \cite{Rab37}, for the
external field of the form%
\begin{equation}
\mathbf{F}=\left(  f_{1}\cos\left(  \Omega t\right)  ,\,f_{2}\sin\left(
\Omega t\right)  ,\,F_{3}\right)  \,, \label{1.3a}%
\end{equation}
where $f_{1,2},\Omega,$ and $F_{3}$ are some constants. For an external field
of the form
\begin{equation}
\mathbf{F}=\left(  F_{1},0,F_{3}\right)  \,,\;F_{1}=\mathrm{const}\,,
\label{1.4}%
\end{equation}
exact solutions for two different functions $F_{3}$ were found in
\cite{BagBaGW01}. These functions are
\begin{equation}
F_{3}=c_{0}\tanh t+c_{1}\,,\;F_{3}=\frac{c_{0}}{\cosh t}+c_{1}\,, \label{1.5a}%
\end{equation}
where $c_{0,1}$ are arbitrary\ real constants. In the work \cite{BagBaGS04},
exact solutions for three sufficiently complicated functions $F_{3}$ were
found. One (the simplest) of these functions reads
\begin{align}
&  F_{3}=c_{0}+\frac{2\left(  c_{1}^{2}-c_{0}^{2}\right)  }{Q+c_{0}%
}\,,\;Q=\left\{
\begin{array}
[c]{c}%
c_{1}\cosh\varphi\,,\;c_{1}^{2}>c_{0}^{2}\,\\
c_{1}\cos\varphi\,,\;c_{1}^{2}<c_{0}^{2}\,
\end{array}
\right.  \,,\nonumber\\
&  \varphi=2\left(  t\sqrt{\left|  c_{1}^{2}-c_{0}^{2}\right|  }+c_{2}\right)
\,, \label{1.6a}%
\end{align}
where $c_{0,1,2}$ are arbitrary real constant.

The aim of the present article is to summarize the properties of the SE, in
order to reveal in detail its relation to other types of equations (associated
equations), and present both a new set of its exact solutions and methods of
generating new solutions.

The article is organized as follows: In sec. 2 we consider associated
equations that are equivalent or (in one way or another) related to the SE. In
secs. 3 and 4 we describe the general solution of the SE and consider the
inverse problem for this equation. A particular, but very important, case of
the self-adjoint SE is considered in sec. 5. We describe here its general
solution, solve the inverse problem, and derive the associated classical
Hamiltonian and Lagrangian systems. In sec. 6 we construct the transformation
matrix that relates different solutions of the SE, and construct the evolution
operator for the SE. The transformation matrix allows one to generate new sets
of exact solution (and the corresponding external fields) from already known
ones. In sec. 7, we represent a set of new exact solutions (together with the
corresponding external fields) of the SE. These solutions are obtained with
the help of a relation (established in sec. 2) between the SE and the
one-dimensional Schr\"{o}dinger equation. In sec. 7, we adapt the Darboux
transformation to a particular case of the SE and thus obtain, as an example,
a new exact solution. The Appendix contains some technical details related to
two-component spinors and associated vectors that are used in the article.

\section{Associated equations}

Below, we present a set of various equations that are equivalent or (in one
way or another) related to the SE. In addition to the definition (\ref{1.1}),
we note the following:

The equation conjugate to the SE has the form%
\begin{equation}
i\dot{V}^{+}=-V^{+}\left(  \mathbf{\sigma F}^{\ast}\right)  \,,\;V^{+}%
=\left(
\begin{array}
[c]{cc}%
v_{1}^{\ast}\left(  t\right)  & v_{2}^{\ast}\left(  t\right)
\end{array}
\right)  \,. \label{1.5}%
\end{equation}

The inner product of two spinors $U$ and $V$ is defined as%
\begin{equation}
\left(  U,V\right)  =U^{+}V=\left(  u_{1}^{\ast}v_{1}+u_{2}^{\ast}%
v_{2}\right)  \,. \label{1.6}%
\end{equation}

The SE and its conjugate equation can be written in the component form as
follows:%
\begin{align}
i\dot{v}_{1}  &  =F_{3}v_{1}+\left(  F_{1}-iF_{2}\right)  v_{2}\,,\;i\dot
{v}_{2}=-F_{3}v_{2}+\left(  F_{1}+iF_{2}\right)  v_{1}\,,\label{1.7}\\
i\dot{v}_{1}^{\ast}  &  =-F_{3}^{\ast}v_{1}^{\ast}-\left(  F_{1}^{\ast}%
+iF_{2}^{\ast}\right)  v_{2}^{\ast}\,,\;i\dot{v}_{2}^{\ast}=F_{3}^{\ast}%
v_{2}^{\ast}-\left(  F_{1}^{\ast}-iF_{2}^{\ast}\right)  v_{1}^{\ast}\,.
\label{1.8}%
\end{align}

The set (\ref{1.8}) can be written in the form of the SE for the spinor
$\bar{V},$
\begin{equation}
\bar{V}=-i\sigma_{2}V^{\ast}=\left(
\begin{array}
[c]{c}%
-v_{2}^{\ast}\\
v_{1}^{\ast}%
\end{array}
\right)  \,, \label{1.9}%
\end{equation}
with an external field $\mathbf{F}^{\ast}$,%
\begin{equation}
i\overset{\cdot}{\bar{V}}=\left(  \mathbf{\sigma F}^{\ast}\right)  \bar{V}\,.
\label{1.10}%
\end{equation}
We refer to the spinor $\bar{V}$ as anti-conjugate to the spinor $V.$

Sometimes, we represent the external field as%
\begin{align}
&  \mathbf{F}=\mathbf{K}+i\mathbf{G}\,,\;\mathbf{K}=\operatorname{Re}%
\mathbf{F\,},\;\mathbf{G}=\operatorname{Im}\mathbf{F}\,,\nonumber\\
&  \mathbf{K}=\left(  K_{k}\right)  \,,\;\mathbf{G}=\left(  G_{k}\right)
\,,\;k=1,2,3\,, \label{1.3}%
\end{align}
where $\mathbf{K}$ and $\mathbf{G}$ are real vectors.

\subsection{Associated Schr\"{o}dinger equations}

\begin{enumerate}
\item  Consider the Schr\"{o}dinger equation in $0+1$ dimensions for a
time-dependent two-component complex spinors $\Psi$. In the general case, the
equation has the form%
\begin{equation}
i\dot{\Psi}=H\Psi\,,\label{ae.1}%
\end{equation}
where the Hamiltonian $H$ is a $2\times2$ complex time-dependent matrix and
$\Psi$ a spinor. The matrix $H$ can always be decomposed in the basic
matrices, $H=hI+\mathbf{\sigma F},$ where $h=h\left(  t\right)  $ and
$\mathbf{F}=\left(  F_{k}\left(  t\right)  ,\;k=1,2,3\right)  .$ Making the
transformation $\Psi=U\exp\left(  -i\int h\,dt\right)  \,,$ we arrive at the
SE for the spinor $U$.

\item  The SE can be reduced to a set of two independent one-dimensional
Schr\"{o}dinger equations with complex potentials in the general case.
Consider the set (\ref{1.7}) for the function $v_{s}\left(  t\right)
,\;s=1,2$. Squaring this set and introducing functions $\psi_{s}\left(
t\right)  ,$%
\begin{equation}
v_{s}=\sqrt{A_{s}}\psi_{s}\,,\;A_{s}=F_{1}+\left(  -1\right)  ^{s}%
iF_{2}\,,\label{3.17}%
\end{equation}
we obtain for the latter functions an independent linear differential equation
of second order:
\begin{align}
&  \ddot{\psi}_{s}-V_{s}\psi_{s}=0\,,\;s=1,2\,,\nonumber\\
&  V_{s}=\frac{3}{4}\left(  \frac{\dot{A}_{s}}{A_{s}}\right)  ^{2}-\frac{1}%
{2}\frac{\ddot{A}_{s}}{A_{s}}-A_{1}A_{2}-F_{3}^{2}-i\left(  -1\right)
^{s}\left(  F_{3}\frac{\dot{A}_{s}}{A_{s}}-\dot{F}_{3}\right)  \,.\label{3.18}%
\end{align}
Each of equations (\ref{3.18}) is a one-dimensional Schr\"{o}dinger equation
with the complex potentials $V_{s}.$
\end{enumerate}

\subsection{Reduction of the external field}

Suppose $V^{\prime}\left(  t\right)  $ is a solution of the SE with an
external field $\mathbf{F}^{\prime}=\left(  F_{1}^{\prime},F_{2}^{\prime
},F_{3}^{\prime}\right)  $. Let us perform a transformation%
\begin{equation}
V=\hat{T}V^{\prime}\,,\;\hat{T}\left(  t\right)  =\exp\left[  i\alpha\left(
\mathbf{\sigma l}\right)  \right]  \,. \label{3.1}%
\end{equation}
Here, $\mathbf{l}$\textbf{ }is an arbitrary constant complex vector, and
$\alpha\left(  t\right)  $ is an arbitrary complex function.

If $\mathbf{l}^{2}\neq0$, then, without loss of generality, we can set
$\mathbf{l}^{2}=1,$ so that the matrix (\ref{3.1}) reads%
\[
\hat{T}=\cos\alpha+i\left(  \mathbf{\sigma l}\right)  \sin\alpha\,.
\]
One can easily verify that the spinor $V\left(  t\right)  $ also obeys an SE
with a reduced external field $\mathbf{F}$ of the form%
\begin{equation}
\mathbf{F}=\left[  \mathbf{F}^{\prime}-\mathbf{l}\left(  \mathbf{F}^{\prime
}\mathbf{l}\right)  \right]  \cos\left(  2\alpha\right)  +\left[
\mathbf{F}^{\prime}\times\mathbf{l}\right]  \sin\left(  2\alpha\right)
+\mathbf{l}\left(  \mathbf{F}^{\prime}\mathbf{l}-\dot{\alpha}\right)  \,.
\label{3.3}%
\end{equation}
Since the transformation matrix $\hat{T}$ is invertible, the SE with the
external field $\mathbf{F}^{\prime}$ and that with the external field
$\mathbf{F},$ given by (\ref{3.3}), are equivalent. If we subject, for
example, the function $\alpha$ to the relation%
\begin{equation}
\dot{\alpha}=\mathbf{F}^{\prime}\mathbf{l}\Longrightarrow\mathbf{Fl}=0\,,
\label{3.4}%
\end{equation}
then the projection of $\mathbf{F}$ onto the direction $\mathbf{l}$\ becomes
zero, i.e., the reduced external field $\mathbf{F}$ has only two nonzero
(complex) components in the plane that is orthogonal to $\mathbf{l}$.

One can also consider complex constant $\mathbf{l}$ with $\mathbf{l}^{2}=0$.
Then%
\[
\hat{T}=1+i\alpha\left(  \mathbf{\sigma l}\right)  \,,
\]
and%
\begin{equation}
\mathbf{F}=\mathbf{F}^{\prime}+2\alpha\left[  \mathbf{F}^{\prime}%
\times\mathbf{l}\right]  +\mathbf{l}\left[  2\alpha^{2}\left(  \mathbf{F}%
^{\prime}\mathbf{l}\right)  -\dot{\alpha}\right]  \,. \label{3.5}%
\end{equation}

By an appropriate choice of a complex $\mathbf{l}$\textbf{,} one can always
reduce to zero any component of both $\mathbf{K}=\operatorname{Re}\mathbf{F}$
and $\mathbf{G}=\operatorname{Im}\mathbf{F}$. However, in this case we cannot
imagine $\mathbf{F}$ as a vector in a fixed plane, unlike in the case of a
real $\mathbf{l}$\textbf{.}

Let us choose the vector $\mathbf{l}$ to be the unit vector in the
$z$-direction, $\mathbf{l}=\left(  0,0,1\right)  ,$ and $\alpha$ to be a
solution of the equation%
\[
\dot{\alpha}=F_{3}^{\prime}\,.
\]
Then the reduced external field $\mathbf{F}$ takes the form%
\begin{equation}
\mathbf{F}=\left(  F_{1},F_{2},0\right)  \,, \label{3.8}%
\end{equation}
where%
\begin{align*}
F_{1}  &  =F_{1}^{\prime}\cos\left(  2\alpha\right)  +F_{2}^{\prime}%
\sin\left(  2\alpha\right)  ,\;F_{2}=F_{2}^{\prime}\cos\left(  2\alpha\right)
-F_{1}^{\prime}\sin\left(  2\alpha\right)  \,,\\
F_{1}^{\prime}  &  =F_{1}\cos\left(  2\alpha\right)  -F_{2}\sin\left(
2\alpha\right)  ,\;F_{2}^{\prime}=F_{2}\cos\left(  2\alpha\right)  +F_{1}%
\sin\left(  2\alpha\right)  \,.
\end{align*}
Choosing $\mathbf{l}=\left(  0,0,1\right)  $ and selecting $\alpha$ to be a
solution of the equations%
\[
F_{1}^{\prime}=F_{1}\cos\left(  2\alpha\right)  \,,\;F_{2}^{\prime}=F_{1}%
\sin\left(  2\alpha\right)  \,,\;F_{3}^{\prime}=F_{3}+\dot{\alpha}\,,
\]
we obtain%
\begin{align}
&  F_{1}=F_{1}^{\prime}\cos\left(  2\alpha\right)  +F_{2}^{\prime}\sin\left(
2\alpha\right)  \,,\;F_{3}=F_{3}^{\prime}-\dot{\alpha}\,,\nonumber\\
&  F_{2}=F_{2}^{\prime}\cos\left(  2\alpha\right)  -F_{1}^{\prime}\sin\left(
2\alpha\right)  =0\,, \label{3.11}%
\end{align}
so that the reduced external field $\mathbf{F}$ takes the form%
\begin{equation}
\mathbf{F}=\left(  F_{1},0,F_{3}\right)  \,. \label{3.10}%
\end{equation}

In addition, one can see that if $V$ is a solution of the SE with the external
field (\ref{3.10}) then:

\begin{enumerate}
\item $U=\left(  2\right)  ^{-1/2}\left(  1+i\sigma_{1}\right)  V\,\ $is a
solution of the SE with the external field%
\begin{equation}
\mathbf{F}=\left(  F_{1},F_{3},0\right)  \,;\label{3.12}%
\end{equation}

\item $U=\sigma_{1}V\,\ $is a solution of the SE with the external field%
\begin{equation}
\mathbf{F}=\left(  F_{1},0,-F_{3}\right)  \,;\label{3.13}%
\end{equation}

\item $U=\sigma_{3}V\,\ $is a solution of the SE with the external field%
\begin{equation}
\mathbf{F}=\left(  -F_{1},0,F_{3}\right)  \,;\label{3.13a}%
\end{equation}

\item $U=\sigma_{2}V\,\ $is a solution of the SE with the external field%
\begin{equation}
\mathbf{F}=\left(  -F_{1},0,-F_{3}\right)  \,;\label{3.13b}%
\end{equation}

\item $U=\left(  2\right)  ^{-1/2}\left(  \sigma_{1}+\sigma_{3}\right)
V\,\ $is a solution of the SE with the external field%
\begin{equation}
\mathbf{F}=\left(  F_{3},0,F_{1}\right)  \,.\label{3.13c}%
\end{equation}
\end{enumerate}

\subsection{Dirac-like equation}

Consider the SE with the external field $\mathbf{F}=\left(  F_{1}%
,0,F_{3}\right)  $. Let us represent the complex spinor $V$ in this SE via two
real spinors $U$ and $W$,%
\begin{equation}
V=U+iW\,.\label{3.25}%
\end{equation}
The SE for $V$ implies the following equations for the real spinors $U$ and
$W$:%
\begin{equation}
\dot{U}=\left(  \mathbf{G}\,\mathbf{\sigma}\right)  U+\left(  \mathbf{K}%
\,\mathbf{\sigma}\right)  W\,,\;\dot{W}=\left(  \mathbf{G}\,\mathbf{\sigma
}\right)  W-\left(  \mathbf{K}\,\mathbf{\sigma}\right)  U\,.\label{3.27}%
\end{equation}
The latter set can be written as%
\begin{equation}
\dot{\Psi}=\left[  \left(  \Sigma\,\mathbf{G}\right)  +\left(  \mathbf{\gamma
}\,\mathbf{K}\right)  \right]  \Psi\,,\label{3.28}%
\end{equation}
where $\Sigma$ and$\;\mathbf{\gamma}$ are Dirac matrices, and $\Psi$ is a
four-component spinor:%
\begin{equation}
\Sigma=\left(
\begin{array}
[c]{cc}%
\mathbf{\sigma} & 0\\
0 & -\mathbf{\sigma}%
\end{array}
\right)  ,\;\mathbf{\gamma}=\left(
\begin{array}
[c]{cc}%
0 & \mathbf{\sigma}\\
-\mathbf{\sigma} & 0
\end{array}
\right)  ,\;\Psi=\left(
\begin{array}
[c]{c}%
U\\
W
\end{array}
\right)  \,.\label{3.29}%
\end{equation}

\subsection{Top equation}

If $V$ is a solution of the SE, then the vectors (\ref{spi.12}) (see Appendix)
obey the following equations:%
\begin{align}
&  \mathbf{\dot{L}}^{v,v}=i\left(  \mathbf{F}^{\ast}-\mathbf{F}\right)
\left(  V,V\right)  +\left[  \left(  \mathbf{F+F}^{\ast}\right)
\times\mathbf{L}^{v,v}\right]  \,,\nonumber\\
&  \mathbf{\dot{L}}^{\bar{v},v}=2\left[  \mathbf{F}\times\mathbf{L}^{\bar
{v},v}\right]  \,,\;\mathbf{\dot{L}}^{v,\bar{v}}=2\left[  \mathbf{F}^{\ast
}\times\mathbf{L}^{v,\bar{v}}\right]  \,. \label{3.14}%
\end{align}
At the same time, the following relations hold:%
\begin{equation}
\mathbf{L}^{v,\dot{v}}=-i\left(  V,V\right)  \mathbf{F}+\left[  \mathbf{F}%
\times\mathbf{L}^{v,v}\right]  \,,\;\mathbf{L}^{\bar{v},\dot{v}}%
=\mathbf{L}^{\overset{\cdot}{\bar{v}},v}=\left[  \mathbf{F}\times
\mathbf{L}^{\bar{v},v}\right]  \,. \label{3.15}%
\end{equation}
In addition, the vectors (\ref{spi.14}) (see Appendix) obey the following
equations:%
\begin{align}
&  \mathbf{\dot{e}}_{1}=2\mathbf{e}_{2}\left(  \mathbf{Kn}\right)
-2\mathbf{n}\left(  \mathbf{Kn}+\mathbf{Ge}_{1}\right)  \,,\nonumber\\
&  \mathbf{\dot{e}}_{2}=2\mathbf{n}\left(  \mathbf{Ke}_{1}-\mathbf{Ge}%
_{2}\right)  -2\mathbf{e}_{1}\left(  \mathbf{Kn}\right)  \,,\nonumber\\
&  \mathbf{\dot{n}}=2\mathbf{e}_{1}\left(  \mathbf{Ke}_{2}+\mathbf{Ge}%
_{1}\right)  +2\mathbf{e}_{2}\left(  \mathbf{Ge}_{2}-\mathbf{Ke}_{1}\right)
\,, \label{3.16}%
\end{align}
with $\mathbf{K}$ and $\mathbf{G}$ given by (\ref{1.3}). Supposing that $V$
obeys the SE, we can find the equations of motion for the parameters
$N,\alpha,\theta,$ and $\varphi$ from the representation (\ref{spi.17}).
Taking into account (\ref{spi.18}) and (\ref{spi.6}), we get
\begin{equation}
2\dot{V}=\left(  2N^{-1}\dot{N}+i\dot{\alpha}-i\dot{\varphi}\cos\theta\right)
V+\left(  \dot{\theta}+i\dot{\varphi}\sin\theta\right)  \exp\left(
i\alpha\right)  \bar{V}\,. \label{3.19}%
\end{equation}
Then, with the help of (\ref{3.19}), (\ref{spi.21}) and (\ref{1.3}), we
finally obtain%
\begin{align}
\dot{\theta}  &  =2\mathbf{Ke}_{\varphi}+2\mathbf{Ge}_{\theta}\,,\;\dot
{\varphi}\sin\theta=2\mathbf{Ge}_{\varphi}-2\mathbf{K\mathbf{e}_{\theta}%
}\,,\label{3.21}\\
\dot{\alpha}  &  =\dot{\varphi}\cos\theta-2\mathbf{Kn}\,,\;\dot{N}%
=N\mathbf{Gn}\,. \label{3.22}%
\end{align}
The set (\ref{3.21}) is autonomous (since it does not contain the functions
$N,\alpha$) and can be written in the compact form%
\begin{equation}
\mathbf{\dot{n}}=2\left[  \mathbf{G-}\left(  \mathbf{Gn}\right)
\mathbf{n}\right]  +2\left[  \mathbf{K}\times\mathbf{n}\right]  \,.
\label{3.24}%
\end{equation}
Thus, the time evolution of the vector $\mathbf{n}$ is determined by the
external field only. If the set (\ref{3.24}) can be integrated to obtain
$\theta\left(  t\right)  $ and $\varphi\left(  t\right)  ,$ then we can find
from (\ref{3.22})%

\begin{equation}
\alpha=\int\left(  \dot{\varphi}\cos\theta-2\mathbf{Kn}\right)  dt\,,\;N=\exp
\left(  \int\mathbf{Gn\,}dt\right)  \,. \label{3.23}%
\end{equation}

Equation (\ref{3.24}) for $\mathbf{G}=0$ is the well-known top equation. It
appears in the gyroscope theory, in the theory of precession of a classical
gyromagnet in a magnetic field, in the theory of electromagnetic resonance
(see \cite{FeyVe57}), and so on. For $\mathbf{G}\neq0,$ this equation
describes a possible damping of the system.

\section{General solution of the spin equation}

The general solution $Y_{\mathrm{gen}}$ of the SE can always be written as%
\begin{equation}
Y_{\mathrm{gen}}=aV+bU\,, \label{4.1a}%
\end{equation}
where $a$ and $b$ are arbitrary complex constants, while $V\left(  t\right)  $
and $U\left(  t\right)  $ are two linearly-independent particular solutions of
the SE. In fact, one needs to know only one particular solution $V,$ since the
other solution $U$ can be constructed from $V$ in quadratures. Indeed,
according to (\ref{spi.6}), one can always present $U$ in the form%
\begin{equation}
U=\alpha V+\beta\bar{V}\,, \label{4.2a}%
\end{equation}
where $\alpha\left(  t\right)  $ and $\beta\left(  t\right)  $\ are complex
functions of time. Substituting (\ref{4.2a}) into the SE, and taking into
account (\ref{1.10}), (\ref{1.3}), we find%
\begin{equation}
\dot{\alpha}V+\dot{\beta}\bar{V}=2\beta\mathbf{\sigma G}\bar{V}\,.
\label{4.3a}%
\end{equation}
Hence,\ multiplying this relation by $V^{+}$ and $\bar{V}^{+}$, we obtain,
with allowance for (\ref{spi.9}) and (\ref{spi.10}),%
\begin{equation}
\left(  V,V\right)  \dot{\beta}=-2\beta\mathbf{L}^{v,v}\mathbf{G}%
\,,\;\dot{\alpha}=2\beta\left(  V,V\right)  ^{-1}\mathbf{L}^{v,\overline{v}%
}\mathbf{G}\,. \label{4.4a}%
\end{equation}
Taking into account (\ref{spi.20}) and (\ref{3.22}), one can rewrite the first
of these equations in the form%
\begin{equation}
N\dot{\beta}=-2\dot{N}\beta\,. \label{4.5a}%
\end{equation}
Equation (\ref{4.5a}) can be easily integrated:%
\begin{equation}
\beta=\beta_{0}N^{-2}=\beta_{0}\left(  V,V\right)  ^{-1}\,, \label{4.6a}%
\end{equation}
where $\beta_{0}$ is an arbitrary complex constant. Then, the second equation
in (\ref{4.4a}) implies%
\begin{equation}
\dot{\alpha}=2\beta_{0}\left(  V,V\right)  ^{-2}\mathbf{L}^{v,\overline{v}%
}\mathbf{G}\,. \label{4.7a}%
\end{equation}
Hence, $\alpha$ can be found by integration:%
\begin{equation}
\alpha=\alpha_{0}+2\beta_{0}\int\left(  V,V\right)  ^{-2}\mathbf{L}%
^{v,\overline{v}}\mathbf{G}dt\,, \label{4.8a}%
\end{equation}
where $\alpha_{0}$ is an arbitrary complex constant. Thus, the general
solution $Y_{\mathrm{gen}}\left(  t\right)  $ of the SE,\ with the known
particular solution $V$ of this equation, has the form%
\begin{equation}
Y_{\mathrm{gen}}=\left[  \alpha_{0}+2\beta_{0}\int\left(  V,V\right)
^{-2}\mathbf{L}^{v,\overline{v}}\mathbf{G}dt\right]  V+\beta_{0}\left(
V,V\right)  ^{-1}\bar{V}\,, \label{4.9a}%
\end{equation}
where $\alpha_{0}$, $\beta_{0}$ are arbitrary complex constants.

\section{Stationary solutions}

Consider the SE with a constant external filed, $\mathbf{F}=\mathrm{const}$.
In this case, we can search for stationary solutions of the form%
\begin{equation}
V\left(  t\right)  =\exp\left(  -i\lambda t\right)  V\,, \label{sts.1}%
\end{equation}
where $V$ is a time-independent spinor subject to the equation%
\begin{equation}
(\mathbf{\sigma F})V=\lambda V\,. \label{sts.2}%
\end{equation}
This equations is analyzed in the Appendix, see below. In particular, for
$\mathbf{F}^{2}\neq0,$ we have two independent solutions $V_{\zeta
}\,,\;\lambda_{\zeta}\,,\;\zeta=\pm1,$%
\begin{align}
&  V_{1}=N_{1}\left(
\begin{array}
[c]{c}%
F_{3}+\sqrt{\mathbf{F}^{2}}\\
iF_{2}+F_{1}%
\end{array}
\right)  ,\;\lambda_{1}=\sqrt{\mathbf{F}^{2}}\,,\nonumber\\
&  V_{-1}=N_{-1}\left(
\begin{array}
[c]{c}%
iF_{2}-F_{1}\\
F_{3}+\sqrt{\mathbf{F}^{2}}%
\end{array}
\right)  ,\;\lambda_{2}=-\sqrt{\mathbf{F}^{2}}\,, \label{sts.3}%
\end{align}
where $N_{\zeta}$ are normalization factors.

\section{The inverse problem for the spin equation}

The inverse problem for the SE can be formulated as follows: provided that a
solution $V$ of the SE is known, is it possible to recover the external field
$\mathbf{F}$ using this solution? The answer to this question can be found in
the general case.

A spinor $V$ gives rise to a triplet of linear vectors (\ref{spi.12}). Let us
decompose the external field in these vectors:%
\begin{equation}
\mathbf{F}=c_{1}\mathbf{L}^{v,v}+c_{2}\mathbf{L}^{v,\overline{v}}%
+c\mathbf{L}^{\overline{v},v}\,, \label{5.1a}%
\end{equation}
where $c_{1},c_{2}\mathbf{,}$ and $c$ are some time-dependent coefficients.
Substituting this expression into the SE and using formula (\ref{spi.11}),
with allowance for (\ref{spi.6}), we find%
\begin{equation}
i\dot{V}=\left(  \mathbf{\sigma F}\right)  V=\left(  V,V\right)  \left(
c_{1}V+2c_{2}\bar{V}\right)  \,. \label{5.2a}%
\end{equation}
Multiplying this relation from the left by $V^{+}$ and $\bar{V}^{+}$, we
obtain%
\begin{equation}
c_{1}=\frac{i\left(  V,\dot{V}\right)  }{\left(  V,V\right)  ^{2}}%
\,,\;c_{2}=\frac{i\left(  \bar{V},\,\dot{V}\right)  }{2\left(  V,V\right)
^{2}}\,. \label{5.3a}%
\end{equation}
Substituting (\ref{5.3a}) into (\ref{5.1a}), we finally have%
\begin{equation}
\mathbf{F}=\frac{i}{2\left(  V,V\right)  ^{2}}\left[  2\left(  V,\dot
{V}\right)  \mathbf{L}^{v,v}+\left(  \bar{V},\dot{V}\right)  \mathbf{L}%
^{v,\overline{v}}\right]  +c\mathbf{L}^{\overline{v},v}\,. \label{5.4a}%
\end{equation}
Here, the complex function $c\left(  t\right)  $ remains arbitrary. Thus,
there exist an infinite number of external fields $\mathbf{F}$ which admit the
same solution of the SE, and the corresponding functional arbitrariness is
completely described.

One can write (\ref{5.4a}) in a different form:%
\begin{equation}
\mathbf{F}=\frac{i}{2\left(  V,V\right)  ^{2}}\left\{  \left(  V,V\right)
\left(  \mathbf{L}^{v,\dot{v}}-\mathbf{L}^{\dot{v},v}\right)  +\left[  \left(
V,\dot{V}\right)  +\left(  \dot{V},V\right)  \right]  \mathbf{L}%
^{v,v}\right\}  +b\mathbf{L}^{\overline{v},v}\,. \label{5.5a}%
\end{equation}
Here, $b\left(  t\right)  $ is also an arbitrary complex function. Expression
(\ref{5.5a}) can be easily reduced to (\ref{5.4a}) with allowance for
(\ref{spi.13}).

The functional arbitrariness arising in the solution of the inverse problem is
related to the fact that the spinor $V$ is given by two complex functions,
whereas the external field $\mathbf{F}$ is defined by three complex functions.

In this way, we have also demonstrated that any complex spinor with an
arbitrary time-dependence (provided that this spinor is differentiable) is a
solution of a certain family of the SE.

Taking into account the explicit form (\ref{spi.17}) of the spinor $V$ and
using formulas (\ref{spi.20}) and (\ref{3.19}), one easily deduces from
(\ref{5.4a}) that%
\begin{equation}
\mathbf{F}=\frac{1}{2}\left[  \left(  \dot{\varphi}\cos\theta-\dot{\alpha
}\right)  \mathbf{n}-\dot{\varphi}\mathbf{e}_{\theta}\sin\theta+\dot{\theta
}\mathbf{e}_{\varphi}\right]  +i\frac{\dot{N}}{N}\mathbf{n+}a\left(
\mathbf{e}_{\theta}+i\mathbf{e}_{\varphi}\right)  \,, \label{5.6a}%
\end{equation}
where $a\left(  t\right)  $ is an arbitrary complex function.

\section{Self-adjoint spin equation}

\subsection{General solution and inverse problem}

We shall refer to the SE as self-adjoint if the external field $\mathbf{F}$ is
real. In this case, according to (\ref{1.3}), we have%
\begin{equation}
\mathbf{F}=\operatorname{Re}\mathbf{F}=\mathbf{K}\,,\;\operatorname{Im}%
\mathbf{F}=\mathbf{G}=0\,. \label{6.1}%
\end{equation}
In this case, the SE has the form of a Schr\"{o}dinger equation,%
\begin{equation}
i\dot{V}=HV\,,\;H=\mathbf{\sigma F}=H^{+}\,, \label{6.2}%
\end{equation}
with a Hermitian Hamiltonian $H.$ Nevertheless, even for a self-adjoint SE,
the one-dimensional Hamiltonians (\ref{3.18}) are not Hermitian in the general case.

Below, we list some properties of a self-adjoint SE, which, generally
speaking, do not take place for a generic complex external field.

The general solution $Y_{\mathrm{gen}}$ of a self-adjoint SE has the form%
\begin{equation}
Y_{\mathrm{gen}}=aV+b\bar{V}\,, \label{6.4}%
\end{equation}
where $V\left(  t\right)  $ is any nonzero particular solution of the SE, and
$a$, $b$ are arbitrary complex constants. This fact follows from (\ref{4.4a}).

For any solution $V$ of a self-adjoint SE, the quantity $N^{2}=\left(
V,V\right)  $ is conserved in time, which is implied by (\ref{3.22}) in case
$\mathbf{G}=0$. However, the reverse is not true. The fact that $N^{2}%
=\mathrm{const}$ does not imply that $V$ is a solution of a self-adjoint SE,
since one can indicate, according to (\ref{5.4a}), a family of complex
external fields of the SE that admit solutions with $N^{2}=\mathrm{const}$.

For an arbitrary nonzero differentiable spinor $V$ subject to the condition
$\left(  V,V\right)  =\mathrm{const}$, there exists only one self-adjoint SE
(only one real external field), whose solution is given by this spinor, and
whose general solution has the form (\ref{6.4}). Indeed, it follows from
(\ref{5.5a}) that in this case a real external field is recovered by the
spinor $V$ in a unique manner:%
\begin{equation}
\mathbf{F}=i\left[  2\left(  V,V\right)  \right]  ^{-1}\left(  \mathbf{L}%
^{v,\dot{v}}-\mathbf{L}^{\dot{v},v}\right)  \,. \label{6.5}%
\end{equation}
It can be easily verified that the same expression for $\mathbf{F}$ arises in
the case when $V$ is replaced by $Y_{\mathrm{gen}}$ from (\ref{6.4}), which
confirms the uniqueness of the external field $\mathbf{F}$. Now, presenting
$V$ in the form (\ref{spi.17}) (see Appendix),\ and setting $N=\mathrm{const}%
$, one obtains a decomposition of $\mathbf{F}$ in the basis vectors of a
spherical coordinate system (\ref{spi.19}):%
\begin{equation}
\mathbf{F}=\frac{1}{2}\left[  \left(  \dot{\varphi}\cos\theta-\dot{\alpha
}\right)  \mathbf{n}-\dot{\varphi}\mathbf{e}_{\theta}\sin\theta+\dot{\theta
}\mathbf{e}_{\varphi}\right]  \,. \label{6.6}%
\end{equation}
Hence, one can find the Cartesian components of the external field and
calculate its square:%
\begin{align}
&  \mathbf{F}=\frac{1}{2}(-\dot{\theta}\sin\varphi-\dot{\alpha}\sin\theta
\cos\varphi,\,\dot{\theta}\cos\varphi-\dot{\alpha}\sin\theta\sin\varphi
,\,\dot{\varphi}\mathbf{-}\dot{\alpha}\cos\theta)\,,\nonumber\\
&  F^{2}=\mathbf{F}^{2}=\frac{1}{4}\left(  \dot{\theta}^{2}+\dot{\varphi}%
^{2}+\dot{\alpha}^{2}-2\dot{\alpha}\dot{\varphi}\cos\theta\right)  \,.
\label{6.7}%
\end{align}
The possibility of an unambiguous recovery of the real external field
$\mathbf{F}$ by a given arbitrary spinor $V\left(  t\right)  $ with a constant
norm also signifies the possibility of generating exactly solvable
self-adjoint SE.

For equations that are associated with a self-adjoint SE, one can state some
additional properties. For instance, the evolution equations (\ref{3.14}) for
the linearly-independent vectors (\ref{spi.12}) become coincident, so that the
vectors (\ref{spi.12}) have to be distinguished by an appropriate choice of
initial conditions.

\subsection{Hamiltonian and Lagrangian forms of self-adjoint spin equation}

Consider the set of equations (\ref{3.21}) for real external fields. In this
case, the set can be written as%
\begin{equation}
\dot{\theta}=2\left(  F_{2}\cos\varphi-F_{1}\sin\varphi\right)  \,,\;\dot
{\varphi}\sin\theta=2F_{3}\sin\theta-2\left(  F_{1}\cos\varphi+F_{2}%
\sin\varphi\right)  \cos\theta\,. \label{6.8}%
\end{equation}
Without loss of generality,\ one can always choose%
\begin{equation}
F_{1}=g\left(  t\right)  \cos\left[  2\alpha\left(  t\right)  \right]
\,,\;F_{2}=g\left(  t\right)  \sin\left[  2\alpha\left(  t\right)  \right]
\,, \label{6.9}%
\end{equation}
where $g\left(  t\right)  $ and $\alpha\left(  t\right)  $ are some real
functions of time. Let us replace $\varphi\left(  t\right)  $ in equation
(\ref{6.8}) by a new function $\Phi\left(  t\right)  $, introduced as%
\begin{equation}
\varphi\left(  t\right)  =\Phi\left(  t\right)  +2\alpha\left(  t\right)  \,.
\label{6.10}%
\end{equation}
It is easy to see that the set (\ref{6.8}) transforms to%
\begin{equation}
\dot{\theta}=-2g\sin\Phi\,,\;\dot{\Phi}\sin\theta=2f\sin\theta-2g\cos\Phi
\cos\theta\,, \label{6.11}%
\end{equation}
where%
\begin{equation}
f=F_{3}\left(  t\right)  -\dot{\alpha}\left(  t\right)  \,. \label{6.12}%
\end{equation}
Notice that the replacement (\ref{6.10}) is equivalent to the transformation
(\ref{3.1}), if one chooses $\mathbf{e}=\left(  0,0,1\right)  $ and selects
$\alpha\left(  t\right)  $ such that the external field takes the form
(\ref{3.10}).

If one introduces the coordinate $q$, the conjugate momentum $p$ and the
Hamiltonian $H$ as follows%
\begin{equation}
q=\cos\theta\,,\;p=-\Phi\,,\;H=2g\sqrt{1-q^{2}}\cos p+2qf\,, \label{6.14}%
\end{equation}
then the set (\ref{6.11}) takes the form of one-dimensional Hamilton equations
\cite{FeyVe57,BagBaGW01},%
\begin{equation}
\dot{q}=\frac{\partial H}{\partial p}\,,\;\dot{p}=-\frac{\partial H}{\partial
q}\,. \label{6.13}%
\end{equation}
Making canonical transformations, we can obtain different forms of the
Hamilton equations that are associated with the self-adjoint SE.

It is straightforward to check that (\ref{6.11}) are the Euler--Lagrange
equations for the Lagrange function%
\begin{equation}
\mathcal{L}=\left[  \left(  1-\gamma\right)  \dot{\theta}\Phi-2g\cos
\Phi\right]  \sin\theta+\left[  \gamma\dot{\Phi}-2f\right]  \cos\theta\,,
\label{6.15}%
\end{equation}
where $\gamma$ is an arbitrary real number.

Finally, the set (\ref{6.11}) leads to a second-order equation for the
function $\theta\left(  t\right)  $:%
\begin{equation}
\overset{\cdot\cdot}{\theta}-\frac{\overset{\cdot}{g}}{g}\overset{\cdot
}{\theta}+2f\sqrt{4g^{2}-\dot{\theta}^{2}}-\left(  4g^{2}-\dot{\theta}%
^{2}\right)  \frac{\cos\theta}{\sin\theta}=0\,. \label{6.16}%
\end{equation}
This equation is also the Euler--Lagrange equation for the Lagrange function%
\begin{equation}
\mathcal{L}=\dot{\theta}\arcsin\left(  \dot{\theta}/2g\right)  \sin
\theta+\sqrt{4g^{2}-\dot{\theta}^{2}}\sin\theta+2f\cos\theta\,. \label{6.17}%
\end{equation}
The above Lagrangian implies the following Hamiltonian%
\begin{equation}
H=-2\left[  g\cos\left(  \frac{p}{\sin\theta}\right)  \sin\theta+f\cos
\theta\right]  \,,\;p=\arcsin\left(  \dot{\theta}/2g\right)  \sin
\theta\label{6.18}%
\end{equation}
which, from the canonical equation (\ref{6.13}), gives the equation
(\ref{6.16}).

\section{More about solutions of the spin equation}

\subsection{The transformation matrix}

Suppose that we know a solution $V_{1}$ of the SE with an external field
$\mathbf{F}_{1}$,%
\[
i\dot{V}_{1}=\left(  \mathbf{\sigma F}_{1}\right)  V_{1}\,,
\]
and wish to find such a nonsingular time-dependent matrix $\hat{T}^{21}$ (in
what follows, it is called the transformation matrix) that a spinor $V_{2}$,%
\begin{equation}
V_{2}\left(  t\right)  =\hat{T}^{21}\left(  t\right)  V_{1}\left(  t\right)
\,, \label{7.1}%
\end{equation}
is a solution of the SE with an external field $\mathbf{F}_{2}$,%
\begin{equation}
i\dot{V}_{2}=\left(  \mathbf{\sigma F}_{2}\right)  V_{2}\,. \label{7.2}%
\end{equation}
It is easy to obtain an equation for the transformation matrix:%
\begin{equation}
i\frac{d}{dt}\hat{T}^{21}=\left(  \mathbf{\sigma F}_{2}\right)  \hat{T}%
^{21}-\hat{T}^{21}\left(  \mathbf{\sigma F}_{1}\right)  \,. \label{7.3}%
\end{equation}
Like any $2\times2$ matrix, the matrix $\hat{T}^{21}$ can be written in the
form%
\begin{equation}
\hat{T}^{21}=a_{0}-i\mathbf{\sigma a}\,,\;\mathbf{a}=\left(  a_{1},a_{2}%
,a_{3}\right)  \,, \label{7.4}%
\end{equation}
where $a_{s}\left(  t\right)  ,$ $s=0\,,1\,,2\,,3$ are some complex functions
of $t$. Substituting (\ref{7.4}) into (\ref{7.3}), and using elementary
properties of the Pauli matrices, one obtains the following set of equations
for the functions $a_{s}$:%
\begin{align}
&  \dot{a}_{0}+\mathbf{aF}_{21}=0\,,\;\mathbf{F}_{21}=\mathbf{F}%
_{2}-\mathbf{F}_{1}\,,\nonumber\\
&  \mathbf{\dot{a}}+2\left[  \mathbf{a\times F}_{1}\right]  +\left[
\mathbf{a\times F}_{21}\right]  -a_{0}\mathbf{F}_{21}=0\,. \label{7.6}%
\end{align}
It is easy to find that $\Delta=\det\hat{T}^{21}=a_{0}^{2}+\mathbf{a}^{2}$ is
an integral of motion. Since the matrix $\hat{T}^{21}$ is determined by
(\ref{7.3})\ only with accuracy up to a constant complex multiplier, we
choose, without loss of generality,%
\begin{equation}
\Delta=a_{0}^{2}+\mathbf{a}^{2}=1\,. \label{7.7}%
\end{equation}
For the inverse matrix $\left(  \hat{T}^{21}\right)  ^{-1}$, we obtain%
\begin{equation}
\left(  \hat{T}^{21}\right)  ^{-1}=\Delta^{-1}\left(  a_{0}+i\mathbf{\sigma
a}\right)  =a_{0}+i\mathbf{\sigma a}\,. \label{7.5}%
\end{equation}

Given $\mathbf{F}_{1}$ and $\mathbf{F}_{2}$, equations (\ref{7.6}) are a
linear homogenous (complex) set of four ordinary differential equations of
first order. Solving this set is completely analogous to solving the SE with
the external field $\mathbf{F}_{2}$, so that we do not achieve any
simplification. However, assuming that the external field $\mathbf{F}_{1}$ and
the matrix $\hat{T}^{21}$ are known, we can obtain from (\ref{7.6}) the
external field $\mathbf{F}_{2}$. It turns out that this problem can be easily
solved. To this end, we need to consider two cases:

\begin{enumerate}
\item  Let $a_{0}\neq0$. We introduce a complex vector $\mathbf{q=q}\left(
t\right)  ,$%
\begin{equation}
\mathbf{q}=\mathbf{a/}a_{0}\,, \label{7.8}%
\end{equation}
such that $\mathbf{q}^{2}\neq-1$. Then (\ref{7.7}) implies%
\begin{equation}
a_{0}=\left(  1+\mathbf{q}^{2}\right)  ^{-1/2}\,. \label{7.9}%
\end{equation}
From (\ref{7.6}), one obtains the equation%
\begin{equation}
\mathbf{\dot{q}}-\mathbf{q}\left(  \mathbf{qF}_{21}\right)  +\left[
\mathbf{q\times F}_{21}\right]  +2\left[  \mathbf{q\times F}_{1}\right]
-\mathbf{F}_{21}=0\,. \label{7.10}%
\end{equation}
This equation allows one to find a unique representation for the external
field $\mathbf{F}_{2},$
\begin{equation}
\mathbf{F}_{2}=\frac{\mathbf{\dot{q}+}\left[  \mathbf{q\times\dot{q}}\right]
+2\left[  \mathbf{q\times F}_{1}\right]  +2\mathbf{q}\left(  \mathbf{qF}%
_{1}\right)  -2\mathbf{q}^{2}\mathbf{F}_{1}}{1+\mathbf{q}^{2}}+\mathbf{F}%
_{1}\,. \label{7.11}%
\end{equation}
In this case, the transformation matrix reads%
\begin{equation}
\hat{T}^{21}=\frac{1-i\mathbf{\sigma q}}{\sqrt{1+\mathbf{q}^{2}}}\,,\;\left(
\hat{T}^{21}\right)  ^{-1}=\frac{1+i\mathbf{\sigma q}}{\sqrt{1+\mathbf{q}^{2}%
}}\,. \label{7.12}%
\end{equation}

\item  Let $a_{0}=0$. We introduce a vector $\mathbf{q=q}\left(  t\right)  ,$%
\[
\mathbf{q}=\mathbf{a}\,.
\]
In this case, the condition (\ref{7.7}) implies $\mathbf{q}^{2}=1,$ and we
obtain from (\ref{7.6})%
\begin{equation}
\mathbf{\dot{q}+}\left[  \mathbf{q\times F}_{21}\right]  +2\left[
\mathbf{q\times F}_{1}\right]  =0\,,\;\mathbf{qF}_{21}=0\,. \label{7.13}%
\end{equation}
From (\ref{7.13}), we uniquely recover $\mathbf{F}_{2}$ in the form%
\begin{equation}
\mathbf{F}_{2}=\left[  \mathbf{q\times\dot{q}}\right]  +2\mathbf{q}\left(
\mathbf{qF}_{1}\right)  -\mathbf{F}_{1}\,,\;\mathbf{q}^{2}=1\,. \label{7.14}%
\end{equation}
The transformation matrix now reads%
\begin{equation}
\hat{T}^{21}=-\left(  \hat{T}^{21}\right)  ^{-1}=-i\mathbf{\sigma q}\,.
\label{7.15}%
\end{equation}
\end{enumerate}

Consequently, having an exact solution $V_{0}$ that corresponds to the
external field $\mathbf{F}_{1},$ we can construct a family of external fields
$\mathbf{F}_{2}$ and the corresponding exact solutions (\ref{7.1}),
parametrized by an arbitrary complex time-dependent vector $\mathbf{q}.$

In the particular case of a self-adjoint SE, the above statement remains valid
if one assumes $\mathbf{q}$ to be a real vector. In this case, the
transformation matrix is unitary.

\subsection{Evolution operator for the spin equation}

In the above consideration, let us choose $\mathbf{F}_{1}=0$ and denote
$\mathbf{F}_{2}=\mathbf{F}$,\ $\hat{T}^{21}=\hat{T}$. Then one can select
$V_{1}$ as an arbitrary constant spinor: $V_{1}=V_{0}=\mathrm{const}$. For
$\mathbf{F}_{1}=0$, one deduces from (\ref{7.3}) that the transformation
matrix $\hat{T}$ obeys the equation%
\begin{equation}
i\frac{d\hat{T}}{dt}=\left(  \mathbf{\sigma F}\right)  \hat{T}\,. \label{8.1}%
\end{equation}
If the transformation matrix $\hat{T}$ is known, the evolution operator
$\hat{R},$ being a solution of equation (\ref{8.1}) with the initial condition
$\hat{R}\left(  0\right)  =I$, can be constructed as follows:%
\begin{equation}
\hat{R}\left(  t\right)  =\hat{T}\left(  t\right)  \hat{T}^{-1}\left(
0\right)  \,. \label{8.5}%
\end{equation}
Using the above expressions for the transformation matrix, we can construct
the evolution operator for the SE with any external field $\mathbf{F}$
according to (\ref{8.5}). The answer reads:

\begin{enumerate}
\item  Let us select an arbitrary complex time-dependent vector $\mathbf{q}%
\left(  t\right)  $ ($\mathbf{q}\left(  0\right)  =\mathbf{q}_{0}$) such that
$\mathbf{q}^{2}\neq-1$. Then the SE with the external field%
\begin{equation}
\mathbf{F}=\frac{\mathbf{\dot{q}+}\left[  \mathbf{q\times\dot{q}}\right]
}{1+\mathbf{q}^{2}} \label{8.6}%
\end{equation}
has the evolution operator of the form%
\begin{equation}
\hat{R}=\frac{\left(  1-i\mathbf{\sigma q}\right)  \left(  1+i\mathbf{\sigma
q}_{0}\right)  }{\sqrt{\left(  1+\mathbf{q}^{2}\right)  \left(  1+\mathbf{q}%
_{0}^{2}\right)  }}=\frac{1+\mathbf{qq}_{0}-i\mathbf{\sigma p}}{\sqrt{\left(
1+\mathbf{q}^{2}\right)  \left(  1+\mathbf{q}_{0}^{2}\right)  }}\,,
\label{8.8a}%
\end{equation}
where $\mathbf{p}=\mathbf{q-q}_{0}+\left[  \mathbf{q}_{0}\times\mathbf{q}%
\right]  $.

\item  Let us select an arbitrary complex unit time-dependent vector
$\mathbf{q}\left(  t\right)  $ ($\mathbf{q}\left(  0\right)  =\mathbf{q}_{0}%
$), $\mathbf{q}^{2}=1$. Then the SE with the external field%
\begin{equation}
\mathbf{F}=\left[  \mathbf{q\times\dot{q}}\right]  \,, \label{8.10}%
\end{equation}
has the evolution operator of the form%
\begin{equation}
\hat{R}=\left(  \mathbf{\sigma q}\right)  \left(  \mathbf{\sigma q}%
_{0}\right)  =\mathbf{qq}_{0}+i\mathbf{\sigma}\left[  \mathbf{q\times q}%
_{0}\right]  \,. \label{8.11}%
\end{equation}
\end{enumerate}

In the case of a self-adjoint SE (real external fields), $\mathbf{q}$ is
selected as a real vector, and the operator $\hat{R}$ is unitary.

\section{Exact solutions of the spin equation}

The first remark: Let $V\left(  t\right)  $ be a solution of the SE with a
given external field $\mathbf{F}$. In this equation, we make the following
change of the variable:%
\begin{equation}
t=T\left(  t^{\prime}\right)  \,, \label{9.1}%
\end{equation}
where $t^{\prime}$ is the new real variable ($T$ is a real invertible
function). Then the SE takes the form
\begin{equation}
i\frac{dV^{\prime}\left(  t\right)  }{dt}=(\mathbf{\sigma F}^{\prime}\left(
t\right)  )V^{\prime}\left(  t\right)  \,, \label{9.2}%
\end{equation}
where%
\begin{equation}
\mathbf{F}^{\prime}\left(  t\right)  =\mathbf{F}\left(  T\left(  t\right)
\right)  \dot{T}\,,\;V^{\prime}\left(  t\right)  =V\left(  T\left(  t\right)
\right)  \,. \label{9.3}%
\end{equation}
Consequently, if one knows a solution of the SE with an external field
$\mathbf{F}$, then one knows a solution of the SE with external fields
$\mathbf{F}^{\prime}$, parametrized by an arbitrary function $T.$ In this
sense, all solutions are divided into equivalence classes. Below, we are going
to list only those solutions of the SE that belong to different classes.

The second remark: We have demonstrated that the SE with an arbitrary external
field can be reduced to an equivalent SE with the external field (\ref{3.10})
which has only two nonzero components:%
\begin{equation}
\mathbf{F}=\left(  F_{1},0,F_{3}\right)  \,. \label{9.3a}%
\end{equation}
Below, we are going to list only solutions for such external fields.

The third remark: Let the components $F_{1}$ and $F_{3}$ of the external field
be linearly dependent. Then, without loss of the generality, we can write
\begin{equation}
F_{1}=q\sin\lambda\,,\;F_{3}=q\cos\lambda\,, \label{9.4}%
\end{equation}
where $q\left(  t\right)  $ is an arbitrary function of time, and $\lambda$ is
a complex constant. Let us define the function $\omega\left(  t\right)  $ by
the relations%
\begin{equation}
\dot{\omega}=q\,,\;\omega\left(  0\right)  =0\,. \label{9.5}%
\end{equation}
Then, the evolution operator for the SE with such an external field has the
form%
\begin{equation}
\hat{R}=\cos\omega-i\left(  \sigma_{1}\sin\lambda+\sigma_{3}\cos
\lambda\right)  \sin\omega t\,. \label{9.6}%
\end{equation}

Especially interesting are solutions of the SE that can be written via the
known special functions.

Below, we consider external fields with such nonzero components $F_{1}$ and
$F_{3}$ that obey the following properties: if solutions of the SE are known
for such external fields, then one can construct solutions for the external
fields%
\begin{equation}
\mathbf{F}=(\alpha F_{1}\left(  \varphi\right)  \,,0,\beta F_{3}\left(
\varphi\right)  )\,,\;\varphi=\omega t+\varphi_{0}\,, \label{9.7}%
\end{equation}
where $\alpha,\beta,$ and $\varphi_{0}$ are arbitrary real constants. We have
succeeded in finding 26 pairs of linearly independent functions $F_{1}$ and
$F_{3}$ that conform to this condition. Below, we list such pairs and present
the spinor $u$ being the corresponding exact solution of the SE. We use the
following notation \cite{G-R}:

$F\left(  \alpha,\beta;\gamma;z\right)  $ is the Gauss hypergeometric function;

$\Phi\left(  \alpha,\gamma;z\right)  $ is the degenerate hypergeometric function;

$D_{p}\left(  z\right)  $ are the parabolic cylinder functions;

$\varphi=\omega t+\varphi_{0}\,$;

$\omega$ and $\varphi_{0}$ are real constants;

$a,b,c$ and $\alpha,\beta,\gamma,\lambda,\mu,\nu$ are complex constants.

\subsection{List of exact solutions}

\begin{enumerate}
\item $F_{1}=at\,,\;F_{3}=bt+c/t:$%
\begin{align*}
u  &  =\left(
\begin{array}
[c]{l}%
at^{\gamma+2}e^{-z/2}\Phi\left(  \alpha+1,\gamma+2;z\right) \\
2\left(  i-c\right)  t^{\gamma}e^{-z/2}\Phi\left(  \alpha,\gamma;z\right)
\end{array}
\right)  ,\\
z  &  =it^{2}\sqrt{a^{2}+b^{2}}\,,\;\alpha=\frac{\gamma}{2}\left(  1+\frac
{b}{\sqrt{a^{2}+b^{2}}}\right)  ,\;\gamma=ic\,.
\end{align*}

\item $F_{1}=a/t\,,\;F_{3}=b/t+ct\,:$%
\begin{align*}
u  &  =\left(
\begin{array}
[c]{l}%
-at^{\gamma-1}e^{-z/2}\Phi\left(  \alpha,\gamma;z\right) \\
\left(  \sqrt{a^{2}+b^{2}}+b\right)  t^{\gamma-1}e^{-z/2}\Phi\left(
\alpha+1,\gamma;z\right)
\end{array}
\right)  ,\\
z  &  =ict^{2}\,,\;2\alpha=i\left(  \sqrt{a^{2}+b^{2}}+b\right)
\,,\;\gamma=1+i\sqrt{a^{2}+b^{2}}\,.
\end{align*}

\item $F_{1}=a/t\,,\;F_{3}=b/t+c\,:$%
\begin{align*}
u  &  =\left(
\begin{array}
[c]{l}%
-at^{\left(  \gamma-1\right)  /2}e^{-z/2}\Phi\left(  \alpha,\gamma;z\right) \\
-iat^{\left(  \gamma-1\right)  /2}e^{-z/2}\Phi\left(  1+\alpha,\gamma
;z\right)
\end{array}
\right)  ,\\
z  &  =2ict\,,\;\alpha=i\left(  \sqrt{a^{2}+b^{2}}+b\right)  \,,\;\gamma
=1+2i\sqrt{a^{2}+b^{2}}\,.
\end{align*}

\item $F_{1}=a/\sin2\varphi\,,\;F_{3}=\left(  b\cos2\varphi+c\right)
/\sin2\varphi\,{\LARGE :}$%
\begin{align*}
u  &  =\left(
\begin{array}
[c]{l}%
-az^{\mu}\left(  1-z\right)  ^{\nu}F\left(  \alpha+1,\beta;\gamma;z\right) \\
\left(  -4i\omega\mu+b+c\right)  z^{\mu}\left(  1-z\right)  ^{\nu}F\left(
\alpha,\beta+1;\gamma;z\right)
\end{array}
\right)  ,\\
z  &  =\sin^{2}\varphi\,,\;\mu=\frac{i}{4\omega}\sqrt{a^{2}+\left(
b+c\right)  ^{2}}\,,\;\nu=\frac{i}{4\omega}\sqrt{a^{2}+\left(  b-c\right)
^{2}}\,,\\
\alpha &  =\mu+\nu-ib/2\omega\,,\;\beta=\mu+\nu+ib/2\omega\,,\;\gamma
=1+2\mu\,.
\end{align*}

\item $F_{1}=a\tan\varphi\,,\;F_{3}=b\tan\varphi+c\cot\varphi\,:$%
\begin{align*}
u  &  =\left(
\begin{array}
[c]{l}%
2\left(  c+i\omega\right)  z^{\mu}\left(  1-z\right)  ^{\nu}F\left(
\alpha,\beta;2\mu;z\right) \\
az^{\mu+1}\left(  1-z\right)  ^{\nu}F\left(  \alpha+1,\beta+1;2\mu+2;z\right)
\end{array}
\right)  ,\\
z  &  =\sin^{2}\varphi\,,\;\mu=-\frac{ic}{2\omega}\,,\;\nu=\frac{i}{2\omega
}\sqrt{a^{2}+b^{2}}\,,\\
\lambda &  =\frac{i}{2\omega}\sqrt{a^{2}+\left(  b-c\right)  ^{2}}%
\,,\;\alpha=\nu+\mu+\lambda\,,\;\beta=\nu+\mu-\lambda\,.
\end{align*}

\item $F_{1}=a/\sin\varphi\,,\;F_{3}=b\tan\varphi+c\cot\varphi\,:$%
\begin{align*}
u  &  =\left(
\begin{array}
[c]{l}%
-az^{\mu}\left(  1-z\right)  ^{\nu+1/2}F\left(  \alpha+1,\beta;2\mu+1;z\right)
\\
\left(  \sqrt{a^{2}+c^{2}}+c\right)  z^{\mu}\left(  1-z\right)  ^{\nu}F\left(
\alpha,\beta;2\mu+1;z\right)
\end{array}
\right)  ,\\
\mu &  =\frac{i}{2\omega}\sqrt{a^{2}+c^{2}}\,,\;\nu=-\frac{ib}{2\omega
}\,,\;z=\sin^{2}\varphi\,,\\
\alpha &  =\mu-\frac{ic}{2\omega}\,,\;\beta=\frac{1}{2}+\mu+2\nu+\frac
{ic}{2\omega}\,.
\end{align*}

\item $F_{1}=a/\cos\varphi\,,\;F_{3}=b\tan\varphi+c\,:$%
\begin{align*}
u  &  =\left(
\begin{array}
[c]{l}%
\left(  \omega+2c-2ib\right)  z^{\mu}\left(  1-z\right)  ^{\nu}F\left(
\alpha,\beta;\gamma;z\right) \\
2iaz^{\mu+1/2}\left(  1-z\right)  ^{\nu}F\left(  \alpha,\beta+1;\gamma
+1;z\right)
\end{array}
\right)  ,\\
z  &  =-e^{-2i\varphi}\,,\;\mu=\frac{c-ib}{2\omega}\,,\;\nu=\frac{i}{\omega
}\sqrt{a^{2}+b^{2}}\,,\\
\alpha &  =\frac{1}{2}+\frac{c}{\omega}+\nu\,,\;\beta=\nu-\frac{ib}{\omega
}\,,\;\gamma=\frac{1}{2}+2\mu\,.
\end{align*}

\item $F_{1}=a/\sinh\varphi\,,\;F_{3}=b\tanh\varphi+c\coth\varphi\,:$%
\begin{align*}
u  &  =\left(
\begin{array}
[c]{l}%
-az^{\mu}\left(  1-z\right)  ^{\nu}F\left(  \alpha,\beta;\gamma;z\right) \\
\left(  -2i\omega\mu a+c\right)  z^{\mu}\left(  1-z\right)  ^{\nu+1/2}F\left(
\alpha,\beta+1;\gamma;z\right)
\end{array}
\right)  ,\\
z  &  =\tanh^{2}\varphi\,,\;\mu=\frac{i}{2\omega}\sqrt{a^{2}+c^{2}}%
\,,\;\nu=\frac{i\left(  b+c\right)  }{2\omega}\,,\\
\alpha &  =\frac{1}{2}+\frac{ib}{\omega}+\beta\,,\;\beta=\mu+\frac{ic}%
{2\omega}\,,\;\gamma=2\mu+1\,.
\end{align*}

\item $F_{1}=a/\cosh\varphi\,,\;F_{3}=b\tanh\varphi+c\coth\varphi\,:$%
\begin{align*}
u  &  =\left(
\begin{array}
[c]{l}%
(2c+i\omega)z^{\mu}\left(  1-z\right)  ^{\nu}F\left(  \alpha,\beta
;\gamma;z\right) \\
az^{\mu+1/2}\left(  1-z\right)  ^{\nu+1/2}F\left(  \alpha+1,\beta
+1;\gamma+1;z\right)
\end{array}
\right)  ,\\
z  &  =\tanh^{2}\varphi\,,\;\mu=-\frac{ic}{2\omega}\,,\;\nu=\frac{i\left(
b+c\right)  }{2\omega}\,,\;\lambda=\frac{1}{2\omega}\sqrt{a^{2}-b^{2}}\,,\\
\alpha &  =\frac{ib}{2\omega}+\lambda\,,\;\beta=\frac{ib}{2\omega}%
-\lambda\,,\;\gamma=\frac{1}{2}-\frac{ic}{\omega}\,.
\end{align*}

\item $F_{1}=a/\sinh2\varphi\,,\;F_{3}=\left(  b\cosh2\varphi+c\right)
/\sinh2\varphi\,:$%
\begin{align*}
u  &  =\left(
\begin{array}
[c]{l}%
-az^{\mu}\left(  1-z\right)  ^{\nu}F\left(  \alpha,\beta;\gamma;z\right) \\
\left(  -4i\omega\mu+b+c\right)  z^{\mu}\left(  1-z\right)  ^{\nu+1}F\left(
\alpha+1,\beta+1;\gamma;z\right)
\end{array}
\right)  ,\\
z  &  =\tanh^{2}\varphi\,,\;\mu=\frac{i}{4\omega}\sqrt{a^{2}+\left(
b+c\right)  ^{2}}\,,\;\lambda=\frac{i}{4\omega}\sqrt{a^{2}+\left(  b-c\right)
^{2}}\,,\\
\nu &  =\frac{ib}{2\omega}\,,\;\alpha=\mu+\nu+\lambda\,,\;\beta=\mu
+\nu-\lambda\,\ ,\;\gamma=1+2\mu\,.
\end{align*}

\item $F_{1}=a/\cosh\varphi\,,\;F_{3}=\left(  b\sinh\varphi+c\right)
/\cosh\varphi\,:$%
\begin{align*}
u  &  =\left(
\begin{array}
[c]{l}%
az^{\mu}\left(  1-z\right)  ^{\nu}F\left(  \alpha,\beta;\gamma;z\right) \\
\left(  2\omega\mu-c+ib\right)  z^{\mu}\left(  1-z\right)  ^{\nu+1}F\left(
\alpha+1,\beta+1;\gamma;z\right)
\end{array}
\right)  ,\\
z  &  =\left(  \frac{e^{\varphi}+i}{e^{\varphi}-i}\right)  ^{2}\,,\;\mu
=\frac{1}{2\omega}\sqrt{a^{2}+\left(  c-ib\right)  ^{2}}\,,\;\alpha=\mu
+\nu+\lambda\,,\\
\lambda &  =\frac{1}{2\omega}\sqrt{a^{2}+\left(  c+ib\right)  ^{2}}%
\,,\;\nu=\frac{ib}{\omega}\,,\;\beta=\mu+\nu-\lambda\,,\;\gamma=1+2\mu\,.
\end{align*}

\item $F_{1}=a\tanh\varphi\,,\;F_{3}=b\tanh\varphi+c\coth\varphi{\LARGE \,}:$%
\begin{align*}
u  &  =\left(
\begin{array}
[c]{l}%
2(c+i\omega)z^{\mu}\left(  1-z\right)  ^{\nu}F\left(  \alpha,\beta
;\gamma;z\right) \\
az^{\mu+1}\left(  1-z\right)  ^{\nu}F\left(  \alpha+1,\beta+1;\gamma
+2;z\right)
\end{array}
\right)  ,\\
z  &  =\tanh^{2}\varphi\,,\;\mu=-\frac{ic}{2\omega}\,,\;\nu=\frac{i}{2\omega
}\sqrt{a^{2}+\left(  b+c\right)  ^{2}}\,,\\
\lambda &  =\frac{i}{2\omega}\sqrt{a^{2}+b^{2}}\,,\;\alpha=\mu+\nu
+\lambda\,,\;\beta=\mu+\nu-\lambda\,,\;\gamma=2\mu\,.
\end{align*}

\item $F_{1}=a\coth\varphi\,,\;F_{3}=b\tanh\varphi+c\coth\varphi\,:$%
\begin{align*}
u  &  =\left(
\begin{array}
[c]{l}%
-az^{\mu}\left(  1-z\right)  ^{\nu}F\left(  \alpha+1,\beta;\gamma;z\right) \\
\left(  2\omega\mu+c\right)  z^{\mu}\left(  1-z\right)  ^{\nu}F\left(
\alpha,\beta+1;\gamma;z\right)
\end{array}
\right)  ,\\
z  &  =\tanh^{2}\varphi\,,\;\mu=\frac{i}{2\omega}\sqrt{a^{2}+c^{2}}%
\,,\;\nu=\frac{i}{2\omega}\sqrt{a^{2}+\left(  b+c\right)  ^{2}}\,,\\
\alpha &  =\mu+\nu+\frac{ib}{2\omega}\,,\;\beta=\mu+\nu-\frac{ib}{2\omega
}\,,\;\gamma=1+2\mu\,.
\end{align*}

\item $F_{1}=a/\cosh\varphi\,,\;F_{3}=b\tanh\varphi+c\,:$%
\begin{align*}
u  &  =\left(
\begin{array}
[c]{l}%
(2b+2c-i\omega)z^{\mu}\left(  1-z\right)  ^{\nu}F\left(  \alpha,\beta
;\gamma;z\right) \\
2az^{\mu+1/2}\left(  1-z\right)  ^{\nu+1/2}F\left(  \alpha+1,\beta
+1,\gamma+1;z\right)
\end{array}
\right)  ,\\
z  &  =\frac{1}{2}\left(  1-\tanh\varphi\right)  \,,\;\alpha=\mu+\nu
+\lambda\,,\;\beta=\mu+\nu-\lambda\,,\\
\mu &  =\frac{i\left(  b+c\right)  }{2\omega}\,,\;\nu=\frac{i\left(
b-c\right)  }{2\omega}\,,\;\gamma=1/2+2\mu\,,\;\lambda=\frac{1}{\omega}%
\sqrt{a^{2}-b^{2}}\,.
\end{align*}

\item $F_{1}=a/\sinh\varphi\,,\;F_{3}=b\coth\varphi+c\,:$%
\begin{align*}
u  &  =\left(
\begin{array}
[c]{l}%
-az^{\mu}\left(  1-z\right)  ^{\nu}F\left(  \alpha,\beta;\gamma;z\right) \\
\left(  -i\omega\mu+b\right)  z^{\mu}\left(  1-z\right)  ^{\nu+1/2}F\left(
\alpha,\beta+1;\gamma;z\right)
\end{array}
\right)  ,\\
z  &  =1-e^{-2\varphi}\,,\;\mu=\frac{i}{\omega}\sqrt{a^{2}+b^{2}}%
\,,\;\nu=\frac{i\left(  b+c\right)  }{2\omega}\,,\\
\alpha &  =\frac{1}{2}+\mu+\frac{ic}{\omega}\,,\;\beta=\mu+\frac{ib}{\omega
}\,,\;\gamma=1+2\mu\,.
\end{align*}
\newline 

\item $F_{1}=a,\;F_{3}=bt+c\,:$%
\begin{align*}
u  &  =\left(
\begin{array}
[c]{l}%
2\sqrt{b}D_{\mu}\left(  z\right) \\
\left(  1+i\right)  aD_{\mu-1}\left(  z\right)
\end{array}
\right)  ,\\
z  &  =\frac{1+i}{\sqrt{b}}\left(  bt+c\right)  \,,\;\mu=-\frac{ia^{2}}{2b}\,.
\end{align*}

\item $F_{1}=a,\;F_{3}=b/t+c\,:$%
\begin{align*}
u  &  =\left(
\begin{array}
[c]{l}%
\left(  1-2ib\right)  t^{\gamma}e^{-z/2}\Phi\left(  \alpha,2\gamma;z\right) \\
-iat^{\gamma+1}e^{-z/2}\Phi\left(  \alpha+1,2\gamma+2,z\right)
\end{array}
\right)  ,\\
z  &  =2it\sqrt{a^{2}+c^{2}}\,,\;\gamma=-ib\,,\;\alpha=\gamma\left(
1-\frac{c}{\sqrt{a^{2}+c^{2}}}\right)  \,.
\end{align*}

\item $F_{1}=a,\;F_{3}=b/t+ct\,:$%
\begin{align*}
u  &  =\left(
\begin{array}
[c]{l}%
(2b+i)t^{\gamma-1/2}e^{-z/2}\Phi\left(  \alpha,\gamma;z\right) \\
at^{\gamma+1/2}e^{-z/2}\Phi\left(  \alpha+1,\gamma+1;z\right)
\end{array}
\right)  ,\\
\,z  &  =ict^{2}\,,\;\alpha=\frac{ia^{2}}{4c}\,,\;\gamma=\frac{1}{2}-ib\,.
\end{align*}

\item $F_{1}=a\,,\;F_{3}=\left(  b\cos2\varphi+c\right)  /\sin2\varphi\,:$%
\begin{align*}
u  &  =\left(
\begin{array}
[c]{l}%
(b+c+i\omega)z^{\mu}\left(  1-z\right)  ^{\nu}F\left(  \alpha,\beta
;\gamma;z\right) \\
az^{\mu+1/2}\left(  1-z\right)  ^{\nu+1/2}F\left(  \alpha+1,\beta
+1;\gamma+1;z\right)
\end{array}
\right)  ,\\
z  &  =\sin^{2}\varphi\,,\;\mu=-\frac{i}{4\omega}\left(  b+c\right)
\,,\;\nu=\frac{i}{4\omega}\left(  c-b\right)  \,,\;\gamma=\frac{1}{2}%
+2\mu\,,\\
\alpha &  =\frac{1}{2\omega}\left(  \sqrt{a^{2}-b^{2}}-ib\right)
\,,\;\beta=-\frac{1}{2\omega}\left(  \sqrt{a^{2}-b^{2}}+ib\right)  \,.
\end{align*}

\item $F_{1}=a\,,\;F_{3}=b\tan\varphi+c\cot\varphi\,:$%
\begin{align*}
u  &  =\left(
\begin{array}
[c]{l}%
(2c+i\omega)z^{\mu}\left(  1-z\right)  ^{\nu}F\left(  \alpha,\beta
;\gamma;z\right) \\
az^{\mu+1/2}\left(  1-z\right)  ^{\nu+1/2}F\left(  \alpha+1,\beta
+1;\gamma+1;z\right)
\end{array}
\right)  ,\\
z  &  =\sin^{2}\varphi\,,\;\mu=-\frac{ic}{2\omega}\,,\;\nu=\frac{ib}{2\omega
}\,,\;\lambda=\frac{1}{2\omega}\sqrt{a^{2}-\left(  b-c\right)  ^{2}}\,,\\
\alpha &  =\mu+\nu+\lambda\,,\;\beta=\mu+\nu-\lambda\,,\;\gamma=\frac{1}%
{2}+2\mu\,.
\end{align*}

\item $F_{1}=a\,,\;F_{3}=b\tan\varphi+c\,:$%
\begin{align*}
u  &  =\left(
\begin{array}
[c]{l}%
az^{\mu}\left(  1-z\right)  ^{\nu}F\left(  \alpha,\beta;\gamma;z\right) \\
\left(  2\omega\mu-c+ib\right)  z^{\mu}\left(  1-z\right)  ^{\nu+1}F\left(
\alpha+1,\beta+1;\gamma;z\right)
\end{array}
\right)  ,\\
z  &  =-e^{-2i\varphi}\,,\;\mu=\frac{1}{2\omega}\sqrt{a^{2}+\left(
c-ib\right)  ^{2}}\,,\;\alpha=\mu+\nu+\lambda\,,\\
\nu &  =\frac{ib}{\omega}\,,\;\beta=\mu+\nu-\lambda\,,\;\gamma=1+2\mu
\,,\;\lambda=\frac{1}{2\omega}\sqrt{a^{2}+\left(  c+ib\right)  ^{2}}.
\end{align*}

\item $F_{1}=a\,,\;F_{3}=b\tanh\varphi+c\coth\varphi\,:$%
\begin{align*}
u  &  =\left(
\begin{array}
[c]{l}%
(2c+i\omega)z^{\mu}\left(  1-z\right)  ^{\nu}F\left(  \alpha,\beta
;\gamma;z\right) \\
az^{\mu+1/2}\left(  1-z\right)  ^{\nu}F\left(  \alpha,\beta+1;\gamma
+1;z\right)
\end{array}
\right)  ,\\
z  &  =\tanh^{2}\varphi\,,\;\mu=-\frac{ic}{2\omega}\,,\;\nu=\frac{i}{2\omega
}\sqrt{a^{2}+\left(  b+c\right)  ^{2}}\,,\\
\gamma &  =\frac{1}{2}+2\mu\,,\;\alpha=\gamma+\nu+\frac{i}{2\omega}\left(
b+c\right)  \,,\;\beta=\nu-\frac{i}{2\omega}\left(  b+c\right)  \,.
\end{align*}

\item $F_{1}=a\,,\;F_{3}=\left(  b\cosh2\varphi+c\right)  /\sinh2\varphi\,:$%
\begin{align*}
u  &  =\left(
\begin{array}
[c]{l}%
(b+c+i\omega)z^{\mu}\left(  1-z\right)  ^{\nu}F\left(  \alpha,\beta
;\gamma;z\right) \\
az^{\mu+1/2}\left(  1-z\right)  ^{\nu}F\left(  \alpha,\beta+1;\gamma
+1;z\right)
\end{array}
\right)  ,\\
z  &  =\tanh^{2}\varphi\,,\;\mu=-\frac{i\left(  b+c\right)  }{4\omega}%
\,,\;\nu=\frac{i}{2\omega}\sqrt{a^{2}+b^{2}}\,,\\
\alpha &  =\frac{1}{2}+\nu-\frac{ic}{2\omega}\,,\;\beta=\nu-\frac{ib}{2\omega
}\,,\;\gamma=\frac{1}{2}+2\mu\,.
\end{align*}

\item $F_{1}=a\,,\;F_{3}=\left(  b\sinh\varphi+c\right)  /\cosh\varphi\,:$%
\begin{align*}
u  &  =\left(
\begin{array}
[c]{l}%
(2b+2ic+i\omega)z^{\mu}\left(  1-z\right)  ^{\nu}F\left(  \alpha,\beta
;\gamma;z\right) \\
2az^{\mu+1/2}\left(  1-z\right)  ^{\nu}F\left(  \alpha,\beta+1;\gamma
+1;z\right)
\end{array}
\right)  ,\\
z  &  =\left(  \frac{e^{\varphi}+i}{e^{\varphi}-i}\right)  ^{2}\,,\;\mu
=\frac{c-ib}{2\omega},\;\nu=\frac{i}{\omega}\sqrt{a^{2}+b^{2}}\,,\\
\alpha &  =\frac{1}{2}+\nu+\frac{c}{\omega}\,,\;\beta=\nu-\frac{ib}{\omega
}\,,\;\gamma=\frac{1}{2}+2\mu\,.
\end{align*}

\item $F_{1}=a\,,\;F_{3}=b\tanh\varphi+c\,:$%
\begin{align*}
u  &  =\left(
\begin{array}
[c]{l}%
az^{\mu}\left(  1-z\right)  ^{\nu}F\left(  \alpha+1,\beta;\gamma;z\right) \\
-\left(  i2\omega\mu+b+c\right)  z^{\mu}\left(  1-z\right)  ^{\nu}F\left(
\alpha,\beta+1;\gamma;z\right)
\end{array}
\right)  ,\\
z  &  =\frac{1}{2}\left(  1-\tanh\varphi\right)  \,,\alpha=\mu+\nu+\frac
{ib}{\omega}\,,\;\nu=\frac{i}{2\omega}\sqrt{a^{2}+\left(  b-c\right)  ^{2}%
}\,,\\
\beta &  =\mu+\nu-\frac{ib}{\omega},\;\gamma=1+2\mu\,,\;\mu=\frac{i}{2\omega
}\sqrt{a^{2}+\left(  b+c\right)  ^{2}}\,.
\end{align*}

\item $F_{1}=a\,,\;F_{3}=b\coth\varphi+c\,:$%
\begin{align*}
u  &  =\left(
\begin{array}
[c]{l}%
2\left(  2b+i\omega\right)  z^{\mu}\left(  1-z\right)  ^{\nu}F\left(
\alpha,\beta;\gamma;z\right) \\
az^{\mu+1}\left(  1-z\right)  ^{\nu}F\left(  \alpha+1,\beta+1;\gamma
+2;z\right)
\end{array}
\right)  ,\\
z  &  =1-e^{-2\varphi}\,,\mu=-\frac{ib}{\omega}\,,\;\nu=\frac{i}{2\omega}%
\sqrt{a^{2}+\left(  b+c\right)  ^{2}},\;\alpha=\nu-\frac{ib}{\omega}%
+\lambda\,,\\
\;\beta &  =\nu-\frac{ib}{\omega}-\lambda\,,\;\gamma=-\frac{2ib}{\omega
}\,,\;\lambda=\frac{i}{2\omega}\sqrt{a^{2}+\left(  b-c\right)  ^{2}}\,.
\end{align*}
\end{enumerate}

\section{Darboux transformation for the spin equation}

Consider the SE with the external field%

\begin{equation}
\mathbf{F}_{\varepsilon}=\left(  F_{1},0,F_{3}\right)  \,,\;F_{1}%
=\varepsilon=\mathrm{const},\;F_{3}=F_{3}\left(  t\right)  \,. \label{1}%
\end{equation}
Exact solutions for such external fields are presented in items 16--26 of the
previous section. The SE with external fields of this form appears in various
physical problems \cite{BagBaGS04,BarCo02,Bar00}. We consider here the Darboux
transformation \cite{Dar1882} for the SE with the potentials (\ref{1}). Such a
transformation allows one to generate new exact solution from the known ones.

For the external field (\ref{1}), the SE for the spinor $V_{\varepsilon}$\ can
be written as the eigenvalue problem%
\begin{equation}
\hat{h}V_{\varepsilon}=\varepsilon V_{\varepsilon}\,,\;\hat{h}=i\sigma
_{1}\frac{d}{dt}+\Lambda\,,\;\Lambda=i\sigma_{2}F_{3}\,. \label{2}%
\end{equation}
The idea of the Darboux transformation in this case can be formulated as
follows: Suppose that the spinor $V_{\varepsilon}$ is known for a given
function $F_{3}$ with any complex $\varepsilon.$ If an operator $\hat{L}$
(called the intertwining operator), that obeys the equation%
\begin{equation}
\hat{L}\hat{h}=\hat{h}^{\prime}\hat{L}\,,\hat{h}^{\prime}=i\sigma_{1}\frac
{d}{dt}+\Lambda^{\prime}\,,\;\Lambda^{\prime}=i\sigma_{2}F_{3}^{\prime}
\label{3}%
\end{equation}
for a function $F_{3}^{\prime}\left(  t\right)  $ is known, then the
eigenvalue problem
\begin{equation}
\hat{h}^{\prime}V_{\varepsilon}^{\prime}=\varepsilon V_{\varepsilon}^{\prime}
\label{4}%
\end{equation}
can be solved as%
\begin{equation}
V_{\varepsilon}^{\prime}=\hat{L}V_{\varepsilon}\,. \label{5}%
\end{equation}
If the intertwining operator $\hat{L}$ is chosen as%
\begin{equation}
\hat{L}=\frac{d}{dt}+A\,, \label{6}%
\end{equation}
where $A\left(  t\right)  $ is a time-dependent $n\times n$ matrix, then the
transformation from $V_{\varepsilon}$ to $V_{\varepsilon}^{\prime}$ is called
the Darboux transformation \cite{MatSa91}. There exists a general method of
constructing the intertwining operators $\hat{L}$ (see, for example,
\cite{NiePeS03} and references therein) for the given eigenvalue problem
(\ref{2}). However, for our purposes the direct application of the general
method cannot be useful. The point is that applying this method one may
violate the specific structure of the initial matrix $\Lambda,$ so that the
final matrix $\Lambda^{\prime}$ will not have the specific structure (\ref{3})
with a real function $F_{3},$ whereas we wish to maintain the structure
(\ref{3}) of the matrix $\Lambda^{\prime},$ i.e., the structure (\ref{1}) of
the external field. Thus, the peculiarity of our problem is that the matrices
$\Lambda$ and $\Lambda^{\prime}$ must have the same block structure and the
Darboux transformation must respect these restrictions. The existence of such
transformations is a nontrivial fact, which we are going to verify below.

The intertwining relation (\ref{3}) with the operator $\hat{L}$ in the form
(\ref{6}) and the matrix $\Lambda^{\prime}$ in the form (\ref{3}) leads to the
following relations:%
\begin{align}
&  \sigma_{1}A-A\sigma_{1}+\sigma_{2}\left(  F_{3}^{\prime}-F_{3}\right)
=0\,,\label{7}\\
&  \sigma_{1}\dot{A}+\sigma_{2}AF_{3}^{\prime}-\sigma_{2}\dot{F}_{3}%
-A\sigma_{2}F_{3}=0\,. \label{8}%
\end{align}
Let us choose%
\begin{equation}
A=\alpha+i\left(  F_{3}-\beta\right)  \sigma_{3}\,, \label{9}%
\end{equation}
where $\alpha\left(  t\right)  $ and $\beta\left(  t\right)  $ are certain
functions. Then we obtain for the function $F_{3}^{\prime}$%
\begin{equation}
F_{3}^{\prime}=2\beta-F_{3} \label{10}%
\end{equation}
and the equations%
\begin{equation}
\dot{\alpha}-2\beta\left(  F_{3}-\beta\right)  =0\,,\;\dot{\beta}%
+2\alpha\left(  F_{3}-\beta\right)  =0 \label{11}%
\end{equation}
for the functions $\alpha$ and $\beta.$ It is easy to see that there exists a
first integral of equation (\ref{11}):%
\begin{equation}
\alpha^{2}+\beta^{2}=R^{2}\,,\;R=\mathrm{const\,}, \label{12}%
\end{equation}
where $R$ is a complex constant in the general case. Note that (\ref{12}) is
satisfied if we choose%
\begin{equation}
\alpha=R\cos\mu\,,\ \beta=R\sin\mu\,, \label{13}%
\end{equation}
with $\mu\left(  t\right)  $ being a real function. Substituting (\ref{13})
into (\ref{11}), we obtain for the function $\mu$ a nonlinear differential
equation:%
\begin{equation}
\dot{\mu}=2\left(  R\sin\mu-F_{3}\right)  \,. \label{14}%
\end{equation}
The time derivative in (\ref{6}) can be taken from equation (\ref{2}). Then we
obtain, with allowance for (\ref{9}) and (\ref{10}),%
\begin{equation}
V_{\varepsilon}^{\prime}=\left[  \alpha-i\left(  \varepsilon\sigma_{1}%
+\beta\sigma_{3}\right)  \right]  V_{\varepsilon}\,. \label{15}%
\end{equation}
Thus, we can see that for the SE with the external field (\ref{1}) there
exists a Darboux transformation that does not violate the structure of the
external field. It has the algebraic form (\ref{15}) and is determined by
solutions of equations (\ref{11}), or by equations (\ref{14}). To complete the
construction, one has to represent solutions of the set (\ref{11}), or
(\ref{14}), with the help of the initial solutions. Such a possibility does
exist and is described below.

Let us fix $\varepsilon=\varepsilon_{0}$ and construct the vector%
\begin{equation}
\mathbf{L}=\left(  \bar{V}_{\varepsilon_{0}},\mathbf{\sigma} V_{\varepsilon
_{0}}\right)  \,, \label{16}%
\end{equation}
see (\ref{spi.9}) from the Appendix. According to equation (\ref{3.14}), this
vector obeys the equation
\begin{equation}
\overset{\cdot}{\mathbf{L}}=2\left[  \mathbf{F}_{\varepsilon_{0}}%
\times\mathbf{L}\right]  \,,\;\mathbf{F}_{\varepsilon_{0}}=\left(
\varepsilon_{0},0,F_{3}\right)  \,. \label{17}%
\end{equation}
In addition, equations (\ref{spi.10}) and (\ref{spi.5}) from Appendix imply
that%
\begin{equation}
\mathbf{L}^{2}=0\,. \label{18}%
\end{equation}
We construct solutions $\alpha$ and $\beta$ of the set (\ref{11}), via the
Cartesian components $L_{i}\,,\;i=1,2,3,$ of $\mathbf{L,}$ as follows:%
\begin{equation}
\alpha=-\varepsilon_{0}\frac{L_{2}}{L_{3}}\,,\;\beta=-\varepsilon_{0}%
\frac{L_{1}}{L_{3}}\,. \label{19}%
\end{equation}
Equation (\ref{18}) implies%
\begin{equation}
\alpha^{2}+\beta^{2}=-\varepsilon_{0}^{2}\,. \label{20}%
\end{equation}
Thus, we have expressed solutions of the set (\ref{11}) via solutions of the
initial equations (\ref{2}) at $\varepsilon_{0}=iR$. Substituting (\ref{19})
into (\ref{15}), one can find the final form of the Darboux transformation:%
\begin{equation}
V_{\varepsilon}^{\prime}=N\sigma_{2}\left[  \left(  \mathbf{\sigma L}\right)
L_{3}^{-1}+\frac{\varepsilon}{\varepsilon_{0}}\sigma_{3}\right]
V_{\varepsilon}\,, \label{21}%
\end{equation}
where $N$ is an arbitrary complex number. Using equation (\ref{spi.21}),
(\ref{evp.12}) from the Appendix, we can transform (\ref{21}) to a different
form:%
\begin{equation}
V_{\varepsilon}^{\prime}=N\left[  2\left(  \bar{V}_{\varepsilon}%
,V_{\varepsilon}\right)  L_{3}^{-1}\sigma_{2}V_{\varepsilon_{0}}%
+i\frac{\varepsilon}{\varepsilon_{0}}\sigma_{1}V_{\varepsilon}\right]  \,.
\label{22}%
\end{equation}

For the constructed Darboux transformation, one can check the following properties:

\begin{enumerate}
\item  If $F_{3}$ is real, then the function $F_{3}^{\prime}$ is also real in
case one chooses the functions $\alpha$ and $\beta$ to be real. This choice is
always possible, according to equations (\ref{11}).

\item  If $F_{3}$ is imaginary, then the function $F_{3}^{\prime}$ is also
imaginary in case one chooses $\alpha$ to be real and $\beta$ to be imaginary.
This choice is always possible, according to equations (\ref{11}).
\end{enumerate}

Consider, finally, a simple example of Darboux transformation. Let $F_{3}=f$
be a constant. Then the general solution of the SE can be obtained from
(\ref{9.4})--(\ref{9.6}). It has the form%
\begin{equation}
V_{\varepsilon}\left(  t\right)  =\left(
\begin{array}
[c]{c}%
i\left(  f-\omega\right)  p\exp\left(  i\omega t\right)  -\varepsilon
q\exp\left(  -i\omega t\right) \\
i\varepsilon p\exp\left(  i\omega t\right)  +\left(  f-\omega\right)
q\exp\left(  -i\omega t\right)
\end{array}
\right)  \,,\;\omega^{2}=f^{2}+\varepsilon^{2}\,, \label{23}%
\end{equation}
where $p$ and $q$ are arbitrary complex constants. The functions $\alpha$ and
$\beta$ can be easily found:%
\begin{align}
&  \alpha=-\frac{\dot{Q}}{2\left(  Q-f\right)  }\,,\;\beta=f+\frac{f^{2}%
-R^{2}}{Q-f}\,,\nonumber\\
&  Q=R\cosh\varphi\,,\;\varphi=2\left(  \omega_{0}t+\varphi_{0}\right)
\,,\;\omega_{0}^{2}=R^{2}-f^{2}\,, \label{24}%
\end{align}
where $R$ and $\varphi_{0}$ are arbitrary complex constants. Then the function
$F_{3}^{\prime}$ can be found from (\ref{10}):%
\begin{equation}
F_{3}^{\prime}=f+2\frac{f^{2}-R^{2}}{Q-f}\,. \label{25}%
\end{equation}

If $f$ is real, then $F_{3}^{\prime}$ is also real in case we choose a real
$R$ and a real $\varphi_{0}$ for $R^{2}>f_{0}^{2}.$ For $R^{2}<f^{2},$
replacing $\varphi_{0}$ by $i\varphi_{0},$ we also obtain a real
$F_{3}^{\prime}$, which is determined by (\ref{25}) with $Q=R\cos\varphi$ and
$\omega_{0}=\sqrt{\left|  R^{2}-f^{2}\right|  }.$ If $f$ is imaginary, then
$F_{3}$ is also imaginary in case we choose an imaginary $R$ .

New solutions $V_{\varepsilon}^{\prime}$ of the SE can be easily constructed
according to formula (\ref{15}). We do not present here their explicit form,
which is quite cumbersome. Note that for any $f\neq0$ such solutions do not
coincide with the new solutions presented in the previous section.

\section{Appendix}

\subsection{Spinors and related vectors}

The following relations hold for any spinor $V$ and its anticonjugate spinor
$\bar{V}$ (\ref{1.9}):%
\begin{equation}
\overline{\left(  \bar{V}\right)  }=-V\,,\;\left(  \bar{V},\bar{V}\right)
=\left(  V,V\right)  \,,\;\left(  \bar{V},V\right)  =\left(  V,\bar{V}\right)
=0\,. \label{spi.5}%
\end{equation}
Since $V\neq0$ and $\bar{V}$ are orthogonal, they are linearly independent.
Therefore, any spinor $U$ can be represented as%
\begin{equation}
U=\left(  V,V\right)  ^{-1}\left[  \left(  V,U\right)  V+\left(  \bar
{V},U\right)  \bar{V}\right]  \,. \label{spi.6}%
\end{equation}
In fact, this means that the completeness relation%
\begin{equation}
VV^{+}+\bar{V}\,\bar{V}^{+}=(V,V)I \label{spi.7}%
\end{equation}
takes place. We note that
\begin{equation}
VU^{+}=\left(
\begin{array}
[c]{cc}%
\upsilon_{1}u_{1}^{\ast} & \upsilon_{1}u_{2}^{\ast}\\
\upsilon_{2}u_{1}^{\ast} & \upsilon_{2}u_{2}^{\ast}%
\end{array}
\right)  ,\,\,\det\left(  VU^{+}\right)  =0\,. \label{spi.3}%
\end{equation}

For any two spinors $U$ and $V$, the following relations hold:%
\begin{align}
&  \,(U,\bar{V})=-(V,\bar{U}),\,(\bar{U},V)=-(\bar{V},U),\,(\bar{U},\bar
{V})=(V,U)\,,\nonumber\\
&  \,(U,\bar{V})(\bar{V},U)=(U,U)(V,V)-(U,V)(V,U)=(\bar{U},V)(V,\bar{U}%
)\geq0\,. \label{spi.8}%
\end{align}

For any two spinors $U$ and $V$, we define a complex vector $\mathbf{L}^{u,v}$
as follows:%
\begin{equation}
\mathbf{L}^{u,v}=\left(  U,\mathbf{\sigma} V\right)  =\left(  u_{1}^{\ast
}v_{2}+u_{2}^{\ast}v_{1},\,iu_{2}^{\ast}v_{1}-iu_{1}^{\ast}v_{2},\,u_{1}%
^{\ast}v_{1}-u_{2}^{\ast}v_{2}\right)  \,. \label{spi.9}%
\end{equation}
The following properties hold:%
\begin{align}
&  \mathrm{i)\,}\left(  \mathbf{L}^{u,v}\right)  ^{\ast}=\mathbf{L}%
^{v,u}\,,\;\mathbf{L}^{\bar{u},\bar{v}}=-\mathbf{L}^{v,u}\,,\nonumber\\
&  \mathrm{ii)\,}\mathbf{L}^{u,v}\mathbf{L}^{u^{\prime},v^{\prime}}=2\left(
U,V^{\prime}\right)  \left(  U^{\prime},V\right)  -\left(  U,V\right)  \left(
U^{\prime},V^{\prime}\right)  \,,\nonumber\\
&  \mathrm{iii)\,}\mathbf{L}^{v,v}\mathbf{L}^{v,v}=\left(  V,V\right)
^{2}\,,\;\mathbf{L}^{\bar{v},v}\mathbf{L}^{\bar{v},v}=\mathbf{L}^{v,\bar{v}%
}\mathbf{L}^{v,\bar{v}}=0\,,\nonumber\\
&  \mathrm{iv)\,}\mathbf{L}^{\bar{v},v}\mathbf{L}^{v,\bar{v}}=2\left(
V,V\right)  ^{2}\,,\;\mathbf{L}^{v,v}\mathbf{L}^{\bar{v},v}=\mathbf{L}%
^{v,v}\mathbf{L}^{v,\bar{v}}=0\,,\nonumber\\
&  \mathrm{v)\,}\left[  \mathbf{L}^{v,\bar{v}}\times\mathbf{L}^{\bar{v}%
,v}\right]  =2i\left(  V,V\right)  \mathbf{L}^{v,v}\,,\;\left[  \mathbf{L}%
^{\bar{v},v}\times\mathbf{L}^{v,v}\right]  =i\left(  V,V\right)
\mathbf{L}^{\bar{v},v}\,,\nonumber\\
&  \left[  \mathbf{L}^{v,v}\times\mathbf{L}^{v,\bar{v}}\right]  =i\left(
V,V\right)  \mathbf{L}^{v,\bar{v}}\,,\nonumber\\
&  \mathrm{vii)\,}\mathbf{L}^{u,v}=\left(  V,V\right)  ^{-1}\left[  \left(
U,V\right)  \mathbf{L}^{v,v}+\left(  U,\bar{V}\right)  \mathbf{L}^{\bar{v}%
,v}\right]  \,. \label{spi.10}%
\end{align}

For any vector $\mathbf{p}$ and any spinor $V,$ we derive, with the help of
(\ref{spi.6}),%
\begin{equation}
\left(  \mathbf{\sigma p}\right)  V=\left(  V,V\right)  ^{-1}\left[  \left(
\mathbf{L}^{v,v}\mathbf{p}\right)  V+\left(  \mathbf{L}^{\bar{v},v}%
\mathbf{p}\right)  \bar{V}\right]  \,. \label{spi.11}%
\end{equation}

Relations (\ref{spi.10}) imply that any spinor $V$ produces three linearly
independent vectors:%
\begin{equation}
\mathbf{L}^{v,v},\;\mathbf{L}^{\bar{v},v},\,\mathbf{L}^{v,\bar{v}}\,.
\label{spi.12}%
\end{equation}
Any complex vector $\mathbf{a}$ can be decomposed in these vectors:%
\begin{align}
&  \mathbf{a}=a_{1}\mathbf{L}^{v,v}+a_{2}\mathbf{L}^{\bar{v},v}+a_{3}%
\mathbf{L}^{v,\bar{v}}\,,\nonumber\\
&  a_{1}=\frac{\mathbf{aL}^{v,v}}{\left(  V,V\right)  ^{2}}\,,\;a_{2}%
=\frac{\mathbf{aL}^{v,\bar{v}}}{2\left(  V,V\right)  ^{2}}\,,\;a_{3}%
=\frac{\mathbf{aL}^{\bar{v},v}}{2\left(  V,V\right)  ^{2}}\,. \label{spi.13}%
\end{align}

It is useful to define three real orthogonal unit vectors $\mathbf{e}%
_{i},\;i=1,2,3,$ with the help of the vectors (\ref{spi.12}),%
\begin{equation}
\mathbf{e}_{1}=\frac{\mathbf{L}^{v,\bar{v}}+\mathbf{L}^{\bar{v},v}}{2\left(
V,V\right)  }\,,\;\mathbf{e}_{2}=i\frac{\mathbf{L}^{v,\bar{v}}-\mathbf{L}%
^{\bar{v},v}}{2\left(  V,V\right)  }\,,\;\mathbf{n}=\frac{\mathbf{L}^{v,v}%
}{\left(  V,V\right)  }\,. \label{spi.14}%
\end{equation}
The latter vectors obey the relations%
\begin{equation}
\mathbf{e}_{i}\mathbf{e}_{j}=\delta_{ij}\,,\;\left[  \mathbf{e}_{i}%
\times\mathbf{e}_{j}\right]  =\epsilon_{ijk}\mathbf{e}_{k}\,,\;\mathbf{e}%
_{3}=\mathbf{n\,,} \label{spi.15}%
\end{equation}
where $\epsilon_{ijk}$ is the Levi-Civita symbol ($\epsilon_{123}=1$). The
inverse relations have the form%
\begin{equation}
\mathbf{L}^{v,v}=\left(  V,V\right)  \mathbf{n}\,,\;\mathbf{L}^{\bar{v}%
,v}=\left(  V,V\right)  \left(  \mathbf{e}_{1}+i\mathbf{e}_{2}\right)
\,,\;\mathbf{L}^{v,\bar{v}}=\left(  V,V\right)  \left(  \mathbf{e}%
_{1}-i\mathbf{e}_{2}\right)  \,. \label{spi.16}%
\end{equation}

Any spinor $V$ can always be represented as%
\begin{equation}
V=Ne^{i\frac{\alpha}{2}}\left(
\begin{array}
[c]{c}%
e^{-i\frac{\varphi}{2}}\cos\frac{\theta}{2}\\
e^{i\frac{\varphi}{2}}\sin\frac{\theta}{2}%
\end{array}
\right)  ,\;N^{2}=\left(  V,V\right)  \,, \label{spi.17}%
\end{equation}
where $N,\alpha,\theta,$ and $\varphi$ are real numbers. The anticonjugate
spinor reads%
\begin{equation}
\bar{V}=Ne^{-i\frac{\alpha}{2}}\left(
\begin{array}
[c]{c}%
-e^{-i\frac{\varphi}{2}}\sin\frac{\theta}{2}\\
e^{i\frac{\varphi}{2}}\cos\frac{\theta}{2}%
\end{array}
\right)  \,. \label{spi.18}%
\end{equation}

Considering $\theta$ and $\varphi$ to be the angles of a spherical reference
frame, we define the corresponding unit orthogonal vectors $\mathbf{\mathbf{e}%
_{\varphi}},\mathbf{e}_{\theta},$ and $\mathbf{n},$%
\begin{align}
&  \mathbf{e}_{\theta}=\left(  \cos\theta\cos\varphi,\cos\theta\sin
\varphi,-\sin\theta\right)  =\left[  \mathbf{e}_{\varphi}\times\mathbf{n}%
\right]  \,,\nonumber\\
&  \mathbf{e}_{\varphi}=\left(  \sin\varphi,\cos\varphi,0\right)  =\left[
\mathbf{n\times e}_{\theta}\right]  \,,\nonumber\\
&  \mathbf{n}=\left(  \sin\theta\cos\varphi,\sin\theta\sin\varphi,\cos
\theta\right)  =\left[  \mathbf{e}_{\theta}\times\mathbf{e}_{\varphi}\right]
\,\,. \label{spi.19}%
\end{align}
In terms of these vectors, the vectors (\ref{spi.12}) and (\ref{spi.14}) can
be written as%
\begin{align}
&  \mathbf{L}^{v,v}=N^{2}\mathbf{n},\;\mathbf{L}^{\bar{v},v}=N^{2}\left(
\mathbf{e}_{\theta}+i\mathbf{e}_{\varphi}\right)  e^{i\alpha},\,\mathbf{L}%
^{v,\bar{v}}=N^{2}\left(  \mathbf{e}_{\theta}-i\mathbf{e}_{\varphi}\right)
e^{-i\alpha},\nonumber\\
&  \mathbf{e}_{1}=\mathbf{e}_{\theta}\cos\alpha-\mathbf{e}_{\varphi}\sin
\alpha\,,\;\mathbf{e}_{2}=\mathbf{e}_{\theta}\sin\alpha+\mathbf{e}_{\varphi
}\cos\alpha\mathbf{\,.} \label{spi.20}%
\end{align}
In addition, it follows from (\ref{spi.11}) and (\ref{spi.20}) that%
\begin{equation}
\mathbf{\sigma F}V=\mathbf{nF}V+\left(  \mathbf{Fe}_{\theta}+i\mathbf{Fe}%
_{\varphi}\right)  \exp\left(  i\alpha\right)  \bar{V}\,. \label{spi.21}%
\end{equation}

\subsection{Eigenvalue problem}

Let us consider the general eigenvalue problem in the space of two-dimensional
spinors:
\begin{equation}
AV=\Lambda V\,. \label{evp.1}%
\end{equation}
Here, $A$ is a given complex $2\times2$ matrix and $\Lambda$ is an eigenvalue.
For nontrivial solutions, the condition
\begin{equation}
\det(A-\Lambda I)=0 \label{evp.2}%
\end{equation}
must hold. One can easily see that
\begin{equation}
\det(A-\Lambda I)=\Lambda^{2}-\Lambda\,\mathrm{tr}A+\det A\,. \label{evp.3}%
\end{equation}
With allowance for (\ref{evp.3}), equation (\ref{evp.2}) implies that two
nonzero eigenvalues are possible:
\begin{equation}
\Lambda_{\zeta}=\frac{1}{2}\left[  \mathrm{tr}A+\zeta\sqrt{\left(
\mathrm{tr}A\right)  ^{2}-4\det A}\right]  \,,\;\zeta=\pm1\,. \label{evp.4}%
\end{equation}

On the other hand, any $2\times2$ matrix $A$ can be written as%
\begin{equation}
A=a_{0}I+(\mathbf{\sigma}\mathbf{a})=\frac{1}{2}\mathrm{tr}A+(\mathbf{\sigma
}\mathbf{a})\,. \label{evp.5}%
\end{equation}
Then the eigenvalue problem (\ref{evp.1}) can be written in the form
\begin{equation}
(\mathbf{\sigma}\mathbf{a})V=\lambda V\,, \label{evp.6}%
\end{equation}
where $\lambda=\Lambda-\frac{1}{2}\mathrm{tr\,}A$\thinspace.

For $\mathbf{a}^{2}\neq0,$ there exist two nontrivial solutions (since
$\det(\mathbf{\sigma}\mathbf{a})=-\mathbf{a}^{2}$). Multiplying (\ref{evp.6})
by $(\mathbf{\sigma}\mathbf{a})$ from the left, and taking into account that
$(\mathbf{\sigma}\mathbf{a})(\mathbf{\sigma}\mathbf{a})=\mathbf{a}^{2},$ we
obtain
\begin{equation}
\mathbf{a}^{2}V=\lambda^{2}V\Longrightarrow\lambda_{\zeta}=\zeta
\sqrt{\mathbf{a}^{2}},\,\,\zeta=\pm1\,,\;\mathbf{a}^{2}\neq0\,. \label{evp.7}%
\end{equation}
The corresponding eigenvectors $V_{\zeta}$ can be written as
\begin{equation}
V_{1}=N_{1}\left(
\begin{array}
[c]{c}%
a_{3}+\sqrt{\mathbf{a}^{2}}\\
a_{1}+ia_{2}%
\end{array}
\right)  ,\,\,V_{-1}=N_{-1}\left(
\begin{array}
[c]{c}%
ia_{2}-a_{1}\\
a_{3}+\sqrt{\mathbf{a}^{2}}%
\end{array}
\right)  , \label{evp.8}%
\end{equation}
where $N_{\zeta}$ are normalization factors.

For $\mathbf{a}^{2}=0,$ the matrix $(\mathbf{\sigma}\mathbf{a})$ is singular,
and there exists a nontrivial solution $V_{0}$ of equation (\ref{evp.6}) for
$\lambda=0.$ Such a solution reads%
\begin{equation}
V_{0}=N\left(
\begin{array}
[c]{c}%
ia_{2}-a_{1}\\
a_{3}%
\end{array}
\right)  ,\;\mathbf{a}^{2}=0\,. \label{evp.9}%
\end{equation}

In a sense, the inverse eigenvalue problem can be formulated as follows: For
any two linearly independent spinors $U$ and $V,$ there exists a vector
$\mathbf{a}$ such that the eigenvalue problem (\ref{evp.6}) with this vector
admits solutions $V_{\zeta}\,,\;\zeta=\pm1,$ of the form
\begin{equation}
V_{1}=U\,,\;V_{-1}=V\,. \label{evp.10}%
\end{equation}
The vector $\mathbf{a}$ is determined by the spinors $U$ and $V$ with accuracy
up to a multiplier $N,$%
\begin{equation}
\mathbf{a}=N\mathbf{L}^{\overline{u},\upsilon}\,, \label{evp.11}%
\end{equation}
where the vector $\mathbf{L}^{\overline{u},\upsilon}$ is defined by
(\ref{spi.9}). To prove the above statement, we consider the matrix $A^{u,v},$%
\begin{align}
&  A^{u,v}=\mathbf{\sigma L}^{u,v}=2VU^{+}-\left(  U,V\right)  I\nonumber\\
&  \,=\left(
\begin{array}
[c]{cc}%
u_{1}^{\ast}v_{1}-u_{2}^{\ast}v_{2} & 2u_{2}^{\ast}v_{1}\\
2u_{1}^{\ast}v_{2} & u_{2}^{\ast}v_{2}-u_{1}^{\ast}v_{1}%
\end{array}
\right)  \,. \label{evp.12}%
\end{align}
This matrix has the following properties:%
\begin{align}
&  A^{u,v}=\left(  A^{v,u}\right)  ^{+}\,,\;\det A^{u,v}=-\left(  U,V\right)
^{2}\,,\nonumber\\
&  A^{u,v}V=\left(  U,V\right)  V\,,\;A^{u,v}\bar{U}=-\left(  U,V\right)
\bar{U}\,,\nonumber\\
&  U^{+}A^{u,v}=\left(  U,V\right)  U^{+}\,,\;\bar{V}^{+}A^{u,v}=-\left(
U,V\right)  \bar{V}^{+}\,. \label{evp.13}%
\end{align}
Then the proof of the statement follows from (\ref{evp.12}) and (\ref{evp.13}).

If we take the spinors $U$ and $V$ \ to be orthogonal,
\begin{equation}
(U,V)=0\Longleftrightarrow U=\alpha\bar{V}\,,\;\alpha=\mathrm{const\,},
\label{epv.14}%
\end{equation}
then (\ref{evp.11}) is reduced to%
\begin{equation}
\mathbf{a}=N^{\prime}\mathbf{L}^{\upsilon,\upsilon},\;N^{\prime}=-\alpha
^{\ast}N\,, \label{epv.15}%
\end{equation}
see (\ref{spi.5}). Note that the vector $\mathbf{L}^{\upsilon,\upsilon}$ is real.

In particular, the above consideration allows us to conclude that the
eigenvectors $U$ and $V$ of equation (\ref{evp.6}) are orthogonal iff the
vector $\mathbf{a}$ is a product of a complex factor and a real vector. In
such a case, the spinors $V$ and $\bar{V}$ obey the completeness relation
(\ref{spi.7}).

\begin{acknowledgement}
V.G.B thanks FAPESP for support and Nuclear Physics Department of S\~{a}o
Paulo University for hospitality, as well as he thanks Russia President grant
SS-1743.2003.2 and RFBR grant 03-02-17615 for partial support; M.C.B. thanks
FAPESP and D.M.G thanks both FAPESP and CNPq for permanent support.
\end{acknowledgement}

\end{document}